%% file: article_finalv3.tex
\documentclass[twocolumn,showpacs,amsmath,amssymb,prd,showkeys,10pt,nofootinbib,superscriptaddress]{revtex4-1}
\usepackage{amsmath, amsmath, amssymb}

\input{defs}

\usepackage{float}% Include files

\usepackage{graphicx}% Include files
\usepackage{dcolumn}% Align table columns on decimal point
\usepackage{bm}% bold math
\usepackage{hyperref}
\usepackage{color}
\usepackage{enumerate}

\definecolor{Blue}{rgb}{0.3,0.3,0.9}
\definecolor{Red}{rgb}{1,0,0}
\definecolor{Green}{rgb}{0,0.6,0}

\begin{document}
\title{Comparison of primordial tensor power spectra from the deformed algebra and dressed metric approaches in loop quantum cosmology}

\author{Boris Bolliet}
\email{boris.bolliet@ens-lyon.fr}
\affiliation{%
\'Ecole Normale Sup\'erieure de Lyon, 46 All\'ee d'Italie, Lyon 69007, France
}%
\affiliation{%
Laboratoire de Physique Subatomique et de Cosmologie, Universit\'e Grenoble-Alpes, CNRS/IN2P3\\
53,avenue des Martyrs, 38026 Grenoble cedex, France
}%

\author{Julien Grain}
\email{julien.grain@ias.u-psud.fr}
\affiliation{%
CNRS, Orsay, France, F-91405}
\affiliation{%
 Universit\'e Paris-Sud 11, Institut d'Astrophysique Spatiale, UMR8617, Orsay, France, F-91405}

\author{Cl\'ement Stahl}
\email{clement.stahl@icranet.org}
\affiliation{%
 Universit\'e Paris-Sud 11, Institut d'Astrophysique Spatiale, UMR8617, Orsay, France, F-91405}
\affiliation{%
CNRS, Orsay, France, F-91405}

\author{Linda Linsefors}%
 \email{linsefors@lpsc.in2p3.fr}
\affiliation{%
Laboratoire de Physique Subatomique et de Cosmologie, Universit\'e Grenoble-Alpes, CNRS/IN2P3\\
53,avenue des Martyrs, 38026 Grenoble cedex, France
}

\author{Aur\'elien Barrau}%
 \email{Aurelien.Barrau@cern.ch}
\affiliation{%
Laboratoire de Physique Subatomique et de Cosmologie, Universit\'e Grenoble-Alpes, CNRS/IN2P3\\
53,avenue des Martyrs, 38026 Grenoble cedex, France
}

\begin{abstract}

Loop quantum cosmology tries to capture the main ideas of loop quantum gravity and to apply them to the Universe as a whole. Two main approaches within this framework have been considered to date for the study of cosmological perturbations: the dressed metric approach and the deformed algebra approach. They both have advantages and drawbacks. In this article, we accurately compare their predictions. In particular, we compute the associated primordial tensor power spectra. We show -- numerically and analytically -- that the large scale behavior is similar for both approaches and compatible with the usual prediction of general relativity. The small scale behavior is, the other way round, drastically different. Most importantly, we show that in a range of wavenumbers explicitly calculated, both approaches do agree on predictions that, in addition, differ from standard general relativity and do not depend on unknown parameters. These features of the power spectrum at intermediate scales might constitute a universal loop quantum cosmology prediction that can hopefully lead to observational tests and constraints. We also present a complete analytical study of the background evolution for the bouncing universe that can be used for other purposes.

\end{abstract}


\maketitle

%%%%%%%%%%%%%%%%%%%%%%%%%%%%%%%%%
%\newpage
%\begin{widetext}
%\tableofcontents
%\end{widetext}
%\newpage
%%%%%%%%%%%%%%%%%%%%%%%%%%%%%%%%%

\section{Introduction}
\label{sec:int}
Loop Quantum Gravity (LQG) is a consistent theory of quantum pseudo-Riemannian geometry that builds on both Einstein gravity and quantum physics, without requiring any fundamentally new principle (like, {\it e.g.}, extra-dimensions or supersymmetry). Several introductory reviews can be found in \cite{lqg_review}. Loop Quantum Cosmology (LQC) is a symmetry reduced version of LQG (see \cite{lqc_review} for introductions) which accounts for the basic cosmological symmetries. At this stage, a fully rigorous derivation of LQC from the mother theory is not yet available. In fact, LQC  imports the main techniques of LQG in the cosmological sector and uses a  ``LQG-like" quantization procedure.  This so-called polymeric quantization relies on a kinematical Hilbert space that is different from the Wheeler-DeWitt one, and therefore evades the Von Neumann uniqueness theorem. Nonetheless, it  has been shown to be well defined when the diffeomorphism invariance is rigorously imposed \cite{unique}. Since there is no operator
associated with the Ashtekar connection but only with its holonomy, the basic variables of LQC are the holonomy of the Ashtekar connection and the flux of the densitized triad, its conjugate momentum. The main result of LQC is that the Big Bang singularity is removed and replaced by a Big Bounce smooth evolution, so that the total energy density cannot be greater than a critical energy density. Intuitively, for sharply peaked states of the background geometry, the Universe undergoes a quantum tunneling from a classical contracting solution to a classical expanding solution.

At the effective level, LQC can be modeled by two kinds of corrections. The inverse-volume corrections \cite{iv} (or inverse-triad, if one relaxes the isotropy hypothesis) are natural cut-off functions of divergences for factors containing inverse powers of densitized triads, arising because of spatial discreetness. The holonomy correction \cite{holo} is instead associated with higher powers of the intrinsic and extrinsic spatial curvature components, stemming from the appearance of holonomies of the Ashtekar connection. As the status of inverse-volume correction is less clear --in particular because of a fiducial-cell dependance-- we only consider in this article the holonomy corrections.

Even when dealing with holonomy corrections only, there are two main ways of considering the effective theory, leading to a lively debate within the LQC community. This study aims at comparing the predictions for cosmological perturbations of both approaches, setting the initial conditions in the same way (that is at the same time and with the same vacuum), which as not been done to date.

The first approach has been developed in \cite{agullo1,agullo2,agullo3} and is referred to as the {\it dressed metric} approach. It relies on a minisuperspace strategy where the homogeneous and isotropic degrees of freedom as well as the inhomogoneous ones (considered as perturbations) are both quantized. The former quantization follows the loop approach whereas the latter is obtained from a Fock-like procedure on a quantum background. The physical inhomogeneous degrees of freedom are given by the Mukhanov-Sasaki variables derived from the linearized classical constraints. The second order Hamiltonian is promoted to be an operator and the quantization is performed using techniques suitable for the quantization of a test field evolving on a quantum background \cite{asht_lew_2009}. The Hilbert space is just the tensor product of a Hilbert space for the background degrees of freedom, with another one for the perturbed degrees of freedom. In the interaction picture, the Schr\"odinger equation for the perturbations was demonstrated to be formally identical to the Schr\"odinger equation for the quantized perturbations evolving on a classical background but using a {\it dressed} metric that encodes the quantum nature of this background.

The second approach, that we refer to as the {\it deformed algebra},  focuses on the well known problem of the consistency of the effective theory. This basically means that the evolution
produced by the model should be consistent with the theory itself. This  translates into the requirement that the
Poisson bracket between two corrected constraints should be proportional to another constraint. The coefficient of proportionality being a
function of the fundamental variables, which makes the situation slightly more subtle than in usual field theories dealing with
simple structure constants. The key point is that the closure of the algebra should also be considered off-shell
\cite{bojo1}. Interestingly, this closure consistency condition is, after the holonomy correction implementation, basically enough to determine the structure of the quantum Poisson bracket algebra \cite{thomCQG1,thomCQG2,Cailleteau:2012fy}. An essential result is that the spacetime structure eventually becomes Euclidean instead of Lorentzian around the bounce, when the total energy density is larger than half the critical energy density. 
This had been overlooked until spherically symmetric inhomogeneity and cosmological perturbations were studied in an anomaly-free way. Without inhomogeneity, one cannot determine the signature because (i) it is impossible to see the relative sign between temporal and spatial derivatives and (ii) the relevant Poisson bracket trivially equals zero in homogeneous models. The signature change is not a consequence of inhomogeneities, the latter rather being used as a test field. There are hints that in the present context, such an effect could really be interpreted as a deep signature change of  space-time rather than a mere tachyonic instability \cite{barrau_bojo}. 

In this specific study, we do not focus on a specific approach. Both have their advantages and drawbacks. The {\it dressed metric}  approach certainly captures more quantum effects, as it deals with the full wave functions. But it faces a problem. In general relativity (GR), there is in principle an infinite number of dynamical laws, all written with respect to different choices of time coordinates. They are all equivalent one to another because of the symmetries of the classical theory and it is legitimate to pick up an arbitrary choice. In the {\it dressed metric}  approach, one is implicitly making use of several such choices, referred to as a background gauge. The mode dynamic is then written in terms of coordinate-invariant combinations of metric and matter perturbations. Only after these steps, one obtains a specific dynamic for the background variables and perturbations, which is written in a Hamiltonian way. Classically, the resulting dynamic does not depend on the coordinate choice and the procedure is valid. But as some degrees of freedom are quantized here, the equations are modified by quantum corrections of different kinds, and nothing still guarantees that the results do not depend on the arbitrary choices made before (that is, the theory may not be covariant or anomaly-free). What is important is the fact that the classical theory enjoys a strong symmetry which is often used in order to simplify the analysis. When one quantizes or modifies the theory, this symmetry must not be violated, or else one may obtain meaningless (gauge-dependent) results. When the dynamics (including dynamical equations and symmetries) is formulated as a constrained system, one gains access to powerful canonical methods by which the consistency of the theory can be easily analyzed. It is of course possible to use another formalism, but not to ignore the problem of potential violations of crucial symmetries \cite{barrau_bojo}. The {\it deformed algebra} approach  does not suffer from this problem and is certainly more obviously consistent. But it does suffer from other difficulties, namely the shape of the modifications is not strictly speaking entirely determined by the anomaly-free condition, there is a kind of tension with the Hojman, Kuchar and Teitelboim theorem \cite{HKT} making the geometrical interpretation difficult, and the fact that the fields are normalized {\it after} the effective quantum corrections were applied to the background, leading to a kind of possibly artificial ``re-quantization" of the theory.\\

The first part of this article is  devoted to  analytical investigations of the background evolution that were already known but not expressed in such a systematic way. This material will also be very useful for the rest of the study, as the shape of the primordial tensor power spectrum depends mainly on the cosmic history. The second part is devoted to the calculation of the infrared and ultraviolet limits of the primordial tensor power spectrum for sharply peaked states in the dressed metric approach. The third part deals with the same issues in the deformed algebra model. In both cases, the initial conditions are set in the same way, in the contracting phase, in order to make a meaningful comparison. The fourth part shows the results of the numerical computations of the full power spectra and some universal features are underlined. In the conclusion, we outline the main differences and similarities between both approaches before giving some perspectives towards observational tests and constraints.

\section{Background evolution: analytical solutions}
\label{sec:bckg}

In this section we study the background evolution at the effective level. Although this has already been studied (see \cite{lqc_review}), our purpose here is to provide analytic solutions which are accurate approximations of the cosmic history over different regions. Our scope is twofold. First, this can give further insights on the effective dynamics of the background, potentially useful for further investigations. Second, these analytic results are developed in the scope of the
forthcoming investigation of tensor perturbations since their equation
of motion obviously involves background quantities
such as the scale factor and the total energy density. 
Here we focus on the most probable dynamics as in \cite{ashtekar_2009,linsefors1}.

\subsection{Overview of the background dynamics}

The background evolution of the quantum universe is described using
the effective, semiclassical dynamics, as derived in loop quantum
cosmology with holonomy corrections. In this article, the background geometry 
is described by the unperturbed metric tensor $\mathrm{g}=-\mathrm{dt}\otimes\mathrm{dt}+a^{2}\delta_{ij}\mathrm{dx^{i}\otimes dx^{j}}$, where $a$ is the scale factor. Dots denote derivatives
with respect to the cosmic time, $\dot{a}\equiv\frac{\partial a}{\partial t}$, and primes denote derivatives with respect to conformal time, related to the cosmic time by $dt=ad\eta$.
The content of the universe is modelled by a single massive scalar field,
$\phi$, with a  quadratic potential, $V\left(\phi\right)=m^2\phi^2/2$. In order to characterize the field evolution we use two dynamical parameters,
the potential energy parameter, $x$, and the kinetic energy parameter, $y$, defined by 
\begin{equation}
x\equiv\frac{m\phi}{\sqrt{2\qsubrm{\rho}{c}}},\,\,\,\,\,\,\,\,\,\,\,\,\,\,\,\,\,\, y\equiv\frac{\dot{\phi}}{\sqrt{2\qsubrm{\rho}{c}}},\label{eq:xyd}
\end{equation}
where $\qsubrm{\rho}{c}$ is the critical density, {\it i.e.}~the maximum value
of the total energy density that can be express as $\rho=\qsubrm{\rho}{c}\left(x^{2}+y^{2}\right)$.
The modified Friedmann equation, as predicted in LQC from the Hamiltonian constraint and the Hamilton equations, is
\begin{equation}
H^{2}=\frac{8\pi G\rho}{3}\left(1-\frac{\rho}{\qsubrm{\rho}{c}}\right)\label{eq:mfe},
\end{equation}
where $H\equiv\frac{\dot{a}}{a}$ is the Hubble parameter. The Klein-Gordon equation for the scalar field is
\begin{equation}
\ddot{\phi}+3H\dot{\phi}+m^{2}\phi=0.\label{eq:kge}
\end{equation}
Equations \eqref{eq:mfe} and \eqref{eq:kge} are recast into
\begin{eqnarray}
\begin{cases}
\dot{H} & =-8\pi G\qsubrm{\rho}{c}y^{2}\left(1-2x^{2}-2y^{2}\right),\\
\dot{x} & =my,\\
\dot{y} & =-3Hy-mx.
\end{cases}\label{eq:sys}\end{eqnarray}

There are two time scales involved in this system of equations. One is given by $1/m$ and corresponds to the classical evolution of the field. The other time scale is $1/\sqrt{G\qsubrm{\rho}{c}}$ and corresponds to the quantum regime of the evolution. Modulo a numerical factor, relevant for the following calculations, the ratio of these two time-scales is
\be
\Gamma\equiv\frac{m}{\sqrt{24\pi G\qsubrm{\rho}{c}}}.\label{eq:gamma}
\end{equation}
If we assume $\Gamma\ll1$, and start with a negative Hubble parameter (contracting universe),
the background dynamics splits into three subsequent phases:
\begin{enumerate}[(i)]
\item Pre-bounce contracting phase,
\item Bouncing phase,
\item Slow-roll inflation.
\end{enumerate}
In each phase, it is possible to get analytical expressions for all the background variables. Note that the value of the inflaton mass preferred by Cosmic Microwave Background (CMB) observations is  $m\simeq10^{-6}\qsubrm{m}{Pl}$. Furthermore,  calculations of the black hole entropy suggests $\qsubrm{\rho}{c}=0.41\qsubrm{m}{Pl}^4$,  leading to  $\Gamma\simeq2\times10^{-7}$. Therefore, asserting $\Gamma\ll1$ is not a strong assumption at all.

\subsection{Initial conditions}

The initial conditions $\{\qsubrm{a}{0},\qsubrm{x}{0},\qsubrm{y}{0}\}$ are set in the remote past, when $\qsubrm{H}{0}<0$ and 
\begin{equation}
\sqrt{\tfrac{\qsubrm{\rho}{0}}{\qsubrm{\rho}{c}}}\ll\Gamma.
\label{eq:c1}
\end{equation}
The subscript `$0$' means that the variables are evaluated at $t=0$. The condition (\ref{eq:c1}) ensures that initially the dynamic is not dominated by the amplification due to the term `$3H$' in (\ref{eq:kge}). We often use polar coordinates for $x$ and $y$: 
\begin{equation}
\begin{cases}
x\left(t\right)=\sqrt{\frac{\rho(t)}{\qsubrm{\rho}{c}}}\sin\left(mt+\qsubrm{\theta}{0}\right),\\
y\left(t\right)=\sqrt{\frac{\rho(t)}{\qsubrm{\rho}{c}}}\cos\left(mt+\qsubrm{\theta}{0}\right).
\end{cases}\label{eq:x0y0}
\end{equation}
The initial value of the energy density is specified with the two numbers $\alpha$
and $\qsubrm{\theta}{0}$:
\begin{equation}
\sqrt{\tfrac{\qsubrm{\rho}{0}}{\qsubrm{\rho}{c}}}=\frac{\Gamma}{\alpha}\left\{1-\frac{\sin(2\qsubrm{\theta}{0})}{4\alpha}\right\}^{-1}.\label{eq:rho0alpha}
\end{equation}
For a given $\alpha\gg1$, such that (\ref{eq:c1}) is valid,  there is a one-to-one correspondence between the
family of solutions to (\ref{eq:sys}) and the interval $\left\{\qsubrm{\theta}{0}|\,0\leq\qsubrm{\theta}{0}<2\pi\right\} $. The choice for this parametrization is clarified in the next section.

\subsection{The pre-bounce classical contracting phase}

As long as \eqref{eq:c1} holds for $\rho\left(t\right)$, the system \eqref{eq:sys} can be solved analytically. In the third line of \eqref{eq:sys}, the term `$3Hy$' can be neglected, compared to $mx$', as their ratio is of order $\mathcal{O}(1/\alpha)$ initially. Then, $x$ and $y$ behave simply as the phase variables of the harmonic oscillator ({\it{i.e.}}~\eqref{eq:x0y0} with constant amplitude). The solution for $y$ can be injected into the equation for $\dot H$ in \eqref{eq:sys} where one neglects `$-2x^2-2y^2$' in comparison to unity in the bracket. The Hubble parameter is replaced by its expression in terms of the energy density \eqref{eq:mfe} where the correction $\rho/\qsubrm{\rho}{c}\ll1$ is neglected. After these replacements, one is left with a first order differential equation over $\rho(t)$ which can be integrated into
\begin{equation}
\sqrt{\tfrac{\rho(t)}{\qsubrm{\rho}{c}}}=\frac{\Gamma}{\alpha}\left\{1-\frac{1}{2\alpha}\left[mt+\tfrac{1}{2}\sin(2mt+2\qsubrm{\theta}{0})\right]\right\}^{-1}.\label{eq:rhoalpha}
\end{equation}
This solution exhibits an oscillatory behavior due to the sine function in the denominator. The oscillations have a period of order $1/m$, much smaller than the time scale of the growth, $\alpha/m$. Moreover, their amplitude is also smaller than the averaged amplitude $\sqrt{{\rho}/{\qsubrm{\rho}{c}}}$ by a factor $\alpha$. When these small and fast oscillations are neglected, the Hubble parameter can be expressed as
\begin{equation}
H(t)=\qsubrm{H}{0}\left(1+\tfrac{3}{2} \qsubrm{H}{0}t\right)^{-1},\label{eq:hubble}
\end{equation}
where the initial Hubble parameter is $\qsubrm{H}{0}=-m/(3\alpha)$. With the parametrization \eqref{eq:rhoalpha},  solutions with the same $\alpha$ but different $\qsubrm{\theta}{0}$'s are all corresponding to the same averaged behavior (there is only a phase difference between them).
From (\ref{eq:hubble}), the scale factor can be computed as a function of cosmic time, and as a function of conformal time after another integration. As the value of the initial conformal time $\qsubrm{\eta}{0}$ can be set arbitrarily, we choose $\qsubrm{\eta}{0}={2}/({\qsubrm{H}{0}\qsubrm{a}{0}})$. With such a choice, the expression for the scale factor simply reads
\be
a\left(\eta\right)=\qsubrm{\lambda}{0}\eta^{2}\,\,\,\,\mathrm{with}\,\,\,\,\qsubrm{\lambda}{0}\equiv\frac{\qsubrm{a}{0}^{3}\qsubrm{H}{0}^2}{4},\label{eq:aac}
\ee
so that the expression of the comoving Hubble radius during the contracting phase is
\be
aH(\eta)=\frac{2}{\eta}. \label{eq:com}
\ee
This is the same behavior as with a universe filled with dust-like matter.  When $H\simeq-{m}/{3}$, the amplification term `$3H$' 
in \eqref{eq:sys} becomes dominant. It corresponds to the end of the pre-bounce contracting phase and the start of the bouncing phase. The contracting phase ends when $\qsubrm{\rho}{A}=\Gamma^2\qsubrm{\rho}{c}$, so at this stage there is no significant quantum effects.

\subsection{The bouncing phase}
\label{subsec:amp}
Let us define $\qsubrm{t}{A}$, the time such that  $H\left(\qsubrm{t}{A}\right)=-{m}/{3}$. One finds $\qsubrm{t}{A}={2}\left(\alpha-1\right)/m$.  Moreover, at $\qsubrm{t}{A}$, if the small and fast oscillations of the field are neglected,
the fractions of potential and kinetic energy are given by
\be
\qsubrm{x}{A}=\Gamma\sin\qsubrm{\theta}{A}\,\,\,\,\,\,\,\mathrm{and}\,\,\,\,\qsubrm{y}{A}=\Gamma\cos\qsubrm{\theta}{A},
\ee
with $\qsubrm{\theta}{A}\equiv2\left(\alpha-1\right)+\qsubrm{\theta}{0}$.
The Hubble parameter keeps increasing (in modulus) until it reaches a maximum, $\qsubrm{H}{max}\equiv{\sqrt{24\pi G\qsubrm{\rho}{c}}}/{6}$. The inverse of $\qsubrm{H}{max}$ has the dimension of a time and gives an estimate of the time scale of this amplification. As a first analysis, in the second equation of the system \eqref{eq:sys}, the time derivative can be replaced by
a factor $\qsubrm{H}{max}$. Then, we find that the ratio between the fraction
of potential and kinetic energy is of order $\sim6\Gamma$, and therefore very small in comparison to unity. This suggests that at the start of the bouncing phase, the kinetic energy parameter
grows very quickly, while the fraction of potential energy remains
of order $\sim\Gamma$. When the kinetic energy is dominant, the system of equations \eqref{eq:sys}
reduces to
\begin{equation}
\begin{cases}
\dot{y}=\sqrt{24\pi G\qsubrm{\rho}{c}}y^{2}\sqrt{1-y^{2}},\\
\dot{x}=my,
\end{cases}\label{eq:sfa}
\end{equation}
which can be solved analytically. The solutions to (\ref{eq:sfa}) shall be valid as long as the kinetic energy dominates over the potential energy. In particular, they are valid at the bounce when the energy density reaches $\qsubrm{\rho}{c}$,
or equivalently when $y\left(\qsubrm{t}{B}\right)=1$. For the time $\qsubrm{t}{B}$, at which the bounce occurs, one finds $\qsubrm{t}{B}=\qsubrm{t}{A}+\frac{1}{m\left|\cos\qsubrm{\theta}{A}\right|}.$

The fractions of kinetic and potential energy during the bouncing phase can be expressed as
\bse
\begin{eqnarray}
	y\left(t\right)&=&\left[1+24\pi G\qsubrm{\rho}{c}(t-\qsubrm{t}{B})^{2}\right]^{-\frac{1}{2}}, \label{eq:mly} \\
	x\left(t\right)&=&\qsubrm{x}{B}+\varepsilon\Gamma\mathrm{arcsinh}\left(\sqrt{24\pi G\qsubrm{\rho}{c}}(t-\qsubrm{t}{B})\right), \label{eq:mlx}
\end{eqnarray}
\ese
where $\varepsilon\equiv\mathrm{sgn}\left(\cos\qsubrm{\theta}{A}\right)$, and the value of the potential energy parameter at the bounce is given by
\begin{equation}
\qsubrm{x}{B}=\qsubrm{x}{A}-\varepsilon\Gamma\ln\left(\tfrac{1}{2}\Gamma\left|\cos\qsubrm{\theta}{A}\right|\right).\label{eq:xb-1}
\end{equation}
The case $\cos\qsubrm{\theta}{A}\ll1$ may appear  problematic. Actually it corresponds to a different evolution of the background, with a phase of {\it deflation} before the bounce. Here we  focus on cases --statistically much more frequent and therefore relevant for phenomenology \cite{linsefors1}-- where a sufficiently long phase of inflation is achieved.
During the bouncing phase, the Hubble parameter and  the scale factor take on a very simple form. The scale factor is related to the kinetic energy parameter by $a =  \qsubrm{a}{B}|y|^{-\frac{1}{3}}$. Consequently, the expression for the scale factor at $\qsubrm{t}{A}$ is 
\begin{equation}
\qsubrm{a}{A}=\qsubrm{a}{B}\left|\Gamma\cos\qsubrm{\theta}{A}\right|^{-\frac{1}{3}}.\label{eq:aaa}
\end{equation}
Using \eqref{eq:com}, we can find the conformal time $\qsubrm{\eta}{A}$ that corresponds to $\qsubrm{t}{A}$. Then, we can use (\ref{eq:aaa}) and (\ref{eq:aac}) in order to write $\qsubrm{\eta}{A}$ in terms of $\qsubrm{\lambda}{0}$. We get
\be
\qsubrm{\eta}{A}=-\left(\frac{6}{m\qsubrm{\lambda}{0}}\right)^{1/3}.\label{eq:etaawithl0}
\ee
After the bounce, the fraction of
potential energy increases. Meanwhile, the fraction of kinetic energy decreases and eventually
becomes smaller than the fraction of potential
energy. This corresponds to the start of slow-roll inflation. 

\subsection{The classical slow-roll inflation}
The total energy density $\rho=\qsubrm{\rho}{c}\left(x^{2}+y^{2}\right)$, 
with $x\left(t\right)$ and $y\left(t\right)$ given by \eqref{eq:mly} and \eqref{eq:mlx},
reaches a minimum at time $\qsubrm{t}{i}$. According to these analytical expressions the total energy density
increases for $t>\qsubrm{t}{i}$. Obviously, this
is irrelevant in an expanding universe without energy sources: the
total energy density must always decrease.  The time $\qsubrm{t}{i}$
 can be computed analytically
by solving $\dot{\rho}\left(\qsubrm{t}{i}\right)=0$. One gets $\qsubrm{t}{i}=\qsubrm{t}{B}+(f/m)$,
where $f$ is expressed in terms of the Lambert $W$ 
function (defined as the solution to $z=W(z)e^{W(z)}$), and $\qsubrm{x}{B}$ is given by
\begin{equation}
f\equiv\sqrt{\frac{2}{W\left(z\right)}} \,\,\,\, \,\,\,\,\mathrm{with}  \,\,\,\,\,\,\,\,z=\frac{8}{\Gamma^2}\exp\left(\frac{2\left|\qsubrm{x}{B}\right|}{\Gamma}\right). \label{eq:f}
\end{equation}
In general, $f$ is of order $\mathcal{O}(1)$. For instance, when $\cos\qsubrm{\theta}{A}=1$ and $\Gamma=2\times10^{-7}$, one gets $f\simeq0.18$.
 At  $\qsubrm{t}{i}$, the fraction of potential energy is calculated with \eqref{eq:mly}, \eqref{eq:mlx} and \eqref{eq:xb-1}. We find
\be
\qsubrm{x}{i}=\qsubrm{x}{A}-2\varepsilon\Gamma\ln\left(\tfrac{1}{2}\Gamma\sqrt{\tfrac{\left|\cos\qsubrm{\theta}{A}\right|}{f}}\right).\label{eq:xi}
\ee
Shortly after $\qsubrm{t}{i}$ (in a time of order $1/(m\ln\Gamma)$), one can show that the fraction of kinetic energy ends up being almost constant. One then has $\qsubrm{y}{i}\equiv-\varepsilon\Gamma$ and the slow-roll conditions are fulfilled. Actually, for the quadratic potential it is enough to check that $\qsubrm{\epsilon}{H}\equiv-\dot{H}/H^2$ is small in comparison to unity for the slow-roll conditions to be valid. We find 
\be
\qsubrm{\epsilon}{H}=3\left|\frac{\Gamma}{\qsubrm{x}{i}}\right|^2,
\ee
which is generally a small number. For  $\cos\qsubrm{\theta}{A}=1$ and $\Gamma=2\times10^{-7}$ one gets $\qsubrm{\epsilon}{H}\simeq0.003$. Slow-roll inflation can start, the system of equation (\ref{eq:sys}) reduces to
\begin{equation}
\begin{cases}
y=-\varepsilon\Gamma,\\
\dot{x}=my,
\end{cases}\label{eq:sri}
\end{equation}
and the Hubble parameter becomes
\be
H(t)=\qsubrm{H}{i}|1-\varepsilon\tfrac{\Gamma}{\qsubrm{x}{i}} m(t-\qsubrm{t}{i})| \label{eq:hsr}.
\ee
where $\qsubrm{H}{i}=\sqrt{8\pi G\qsubrm{\rho}{c}/3}|\qsubrm{x}{i}|$. We can also use \eqref{eq:hsr} to compute the scale factor at $\qsubrm{t}{i}$ with $|\qsubrm{y}{i}|=\Gamma$. We get
\be
\qsubrm{a}{i}=\qsubrm{a}{B}\Gamma^{-\frac{1}{3}}.\label{eq:ai}
\ee
Note that at the start of slow-roll inflation the total energy density is smaller than the critical energy density by a factor $\Gamma^2$. Therefore, when slow-roll inflation starts,  the universe is already classical (since quantum corrections are negligible). 

We stress that all the analytical approximations derived above have been checked against numerical integrations of equation \eqref{eq:sys}. This has been done for each one of the three subsequent phases as well as for the matching between them.

\section{Power spectrum in the dressed metric approach}
\label{sec:irpow}
\subsection{Preliminaries on the dressed metric approach}
The dressed metric approach for both scalar and tensor cosmological perturbations in LQC has been developed in \cite{agullo1,agullo2,agullo3}. Focusing on the tensor modes, the primordial power spectrum at the end of inflation is defined in terms of the mode functions of the Mukhanov-Sasaki variables, denoted $v_k$,  as\footnote{This model is parity invariant and the two helicity states of the tensor mode are equally amplified. The summation over the helicity states is implicitly done in our definition of the primordial power spectrum.}
\begin{equation}
\qsubrm{\mathcal{P}}{T}(k)=\frac{32 G k^3}{\pi}\left|\frac{v_k(\qsubrm{\eta}{e})}{a(\qsubrm{\eta}{e})}\right|^2,\label{pdef}
\end{equation}
with $\qsubrm{\eta}{e}$ standing for the end of inflation. 

It is worth  mentioning that the precise knowledge of $\qsubrm{\eta}{e}$ is not mandatory for the derivation of the primordial power spectrum in both the infrared (IR) and ultraviolet (UV) limits. For the IR limit, this is because infrared modes are (by definition) mainly amplified during the contraction and the contribution of inflation is suppressed as compared to the previous phases. In the UV, this is because we focus on modes that crossed the horizon during inflation, so that their amplitude has remained constant after a few e-folds.\\

In order to obtain the power spectrum, one has to solve the equation of motion for the mode functions, $v_k(\eta)$, with given initial conditions. In conformal time, this equation takes the form of a Schr\"odinger equation
\begin{equation}
	v''_k(\eta)+\left(k^2-\frac{\left<\tilde{a}^{\prime\prime}\right>}{\left<\tilde{a}\right>}\right)v_k(\eta)=0,\label{eq:eschro}
\end{equation}
where $\tilde{a}$ is a {\it dressed} scale factor and $\left< . \right>$ refers to the quantum expectation value on background states. This takes into account the width of the background wave function and has {\it a priori} no reason to be equal to the scale factor, $a(t)$, solution to the modified Friedmann equation (corresponding to the scale factor traced by the {\it peak} of the sharply peaked wave function). However, it is argued in \cite{agullo3} that for  sharply peaked background states, the dressed effective potential term, $\left<\tilde{a}''\right>/\left<\tilde{a}\right>$, is very well approximated by its peaked value, $a^{\prime\prime}/a$, from the bounce up to the entire expanding phase. We expect this approximation to be valid from the bounce down to the classical contracting phase since this also corresponds to a more and more classical universe when going backward in time from the bounce. With this approximation, (\ref{eq:eschro}) becomes
\begin{equation}
	v''_k(\eta)+\left(k^2-\frac{a''}{a}\right)v_k(\eta)=0, \label{eom}
\end{equation}
where the scale factor is now solution to the modified Friedmann equation, and the analytical results derived in Sec.~\ref{sec:bckg} can be used for the background variables. 

\subsection{Calculation of the IR limit}
\subsubsection{Definition of the IR regime}
The IR limit of the primordial power spectrum is obtained by considering the modes which stopped oscillating with time and were frozen during the pre-bounce contracting phase. The  freezing of a mode happens when its wavenumber becomes smaller than the effective potential, $\sqrt{a''/a}$. With the analytical expressions given in the previous section, one finds that during the contraction, 
\begin{equation}
\frac{a^{\prime\prime}}{a}=\frac{2}{\eta^{2}}.\label{eq:appa}
\end{equation}
Thus, an infrared mode with a wavenumber $k$ crosses the effective potential at a conformal time $\left|\eta_k\right|\equiv\sqrt{2}/k$. Its amplitude is frozen from that time up to the end of inflation, as $k^2$ remains smaller than $a''/a$. Since $-\infty<\eta<\eta_A$ (with $\eta_A<0$), the modes that crossed the potential during the contracting phase are in the range $0<k<\qsubrm{k}{IR}$, with $\qsubrm{k}{IR}$ defined by the mode that crossed the effective potential at the beginning of the bouncing phase. With (\ref{eq:appa}), (\ref{eq:aaa}),  (\ref{eq:aac}) and (\ref{eq:etaawithl0}) we find
\be
\qsubrm{k}{IR} = \frac{\qsubrm{a}{B}}{3\sqrt{2}}\left(\frac{m^2\sqrt{24\pi G\qsubrm{\rho}{c}}}{\left|\cos\qsubrm{\theta}{A}\right|}\right)^{1/3}.
 \label{eq:ka}
\ee
The IR limit stands for the modes such that $k\ll \qsubrm{k}{IR}$. \\

\subsubsection{Primordial power spectrum in the IR regime}
For infrared modes, from the potential crossing $\eta_k$ to the end of inflation $\qsubrm{\eta}{e}$, the solution to the equation of motion (\ref{eom}) is therefore well approximated by
\begin{equation}
	\qsuprm{v}{IR}_k(\eta)=\alpha_ka(\eta)+\beta_ka(\eta)\displaystyle\int^\eta_{\eta_{\star}}\frac{d\eta'}{a^2(\eta')}+\mathcal{O}((k/\qsubrm{k}{IR})^2) \label{approxir},
\end{equation}
where $\alpha_k$ and $\beta_k$ are two constants to be determined. The value of $\eta_{\star}$ can be conveniently set by requiring the term proportional to $\alpha_k$ to be solely decaying, and the term proportional to $\eta_k$ to be solely growing. During the contracting phase, the term proportional to $\alpha_k$ is  clearly decaying since $a(\eta)$ is decreasing. A convenient choice for $\eta_{\star}$ is such that the term proportional to $\beta_k$ must be solely growing. Since $a(\eta)=\qsubrm{\lambda}{0}\eta^2$, the term proportional to $\beta_k$ has a time dependance $\sim\eta^2(\eta^{-3}-\eta_{\star}^{-3})$, in which the part proportional to $\eta^2/\eta_{\star}^3$ is decaying ($\eta<0$) and we send $\eta_{\star}$ to $(-\infty)$ to remove it\footnote{During the contracting phase, the identification of the growing and decaying modes differs from that identification during inflation. Because the Universe is expanding during inflation, the term $\alpha_ka(\eta)$ is solely growing (while it is solely decaying during contraction). Then the term $\beta_k\int^\eta_{\eta_{\star}}d\eta'/a^2(\eta')$ can be made solely decaying in an inflationary universe by setting $\eta_{\star}=\qsubrm{\eta}{e}$ (while it is made solely growing during contraction by setting $\eta_{\star}\to-\infty$).}. 

From \eqref{approxir} and \eqref{pdef}, the expression of the IR limit of the spectrum reads 
\begin{equation}
\qsuprm{\qsubrm{\mathcal{P}}{T}(k)}{IR}=\frac{32 G k^3}{\pi}\left|\alpha_k+\beta_k I(\qsubrm{\eta}{e})\right|^2,\label{pir}
\end{equation}
 where $I(\qsubrm{\eta}{e})$ is the integral defined as
\be
I(\qsubrm{\eta}{e}) \equiv \int^{\qsubrm{\eta}{e}}_{-\infty}\frac{d\eta}{a^2}. \label{eq:int}
\ee
The calculation of the IR limit proceeds in two steps. First, we compute $\alpha_k$ and $\beta_k$ by matching (\ref{approxir}) to a set of solutions defined in the contracting phase. As we shall see, this determines the scale dependence of the primordial power spectrum in the infrared regime. The second step is the calculation $I(\qsubrm{\eta}{e})$ using the analytical solutions for the background, obtained in Sec.~\ref{sec:bckg}. This second step sets the amplitude of the power spectrum. The expression of the primordial spectrum is finally obtained by gathering the expressions of $\alpha_k$, $\beta_k$ and $I(\eta_e)$. \\

In order to derive the expressions of  $\alpha_k$ and $\beta_k$, the approximate solution given in  (\ref{approxir}) (valid in the IR only but from $\eta_k$ to $\qsubrm{\eta}{e}$) has to be matched with a set of solutions  to the equation (\ref{eom}), during the  contracting phase. With $a''/a=2/\eta^2$, this set of solutions corresponds to the linear combinations of the Hankel functions of order $\nu=3/2$:
\begin{equation}
	\qsuprm{v}{C}_k(\eta)=\sqrt{-k\eta}\left[A_k \qsubrm{H}{3/2}(-k\eta)+B_k\qsubrm{H}{3/2}^\star(-k\eta)\right],\label{eq:vc}
\end{equation}
where the superscript `$\mathrm{C}$' recalls that (\ref{eq:vc}) is valid only during the contracting phase. In order to specify $A_k$ and $B_k$ we match (\ref{eq:vc}) with the Minkowski vacuum in the remote past, {\it i.e.}~$v_k(\eta\to-\infty)=e^{-ik\eta}/\sqrt{2k}$. This requirement leads to
\begin{equation}
A_k=\sqrt{\frac{\pi}{4k}}~~~\mathrm{and}~~~B_k=0,
\end{equation}
up to a phase which is has no importance here\footnote{This also fits with the appropriate Wronskian condition as required for the quantization {\it \`a la} Fock of the tensor perturbations field.}. A set of solutions valid in the range $-\infty<\eta<\eta_A$, and corresponding to the Minkowski vacuum, is thus
\begin{equation}
	\qsuprm{v}{C}_k(\eta)=\tfrac{1}{2}\sqrt{-\pi\eta}\qsubrm{H}{3/2}(-k\eta).  \label{approxpast}
\end{equation}
Since $\eta_k\ll\qsubrm{\eta}{A}$ for infrared modes, the IR limit of  (\ref{approxpast}) has to coincide with (\ref{approxir}) in the interval $\eta_k\lesssim \eta\lesssim \qsubrm{\eta}{A}$.  At a given $\eta$ in this interval, we calculate the asymptotic limit of the Hankel function when $k\to0$. This leads to
\begin{equation}
	\lim_{k\to0}\qsuprm{v}{C}_k(\eta)=\frac{i}{\sqrt{2}k^{3/2}\eta} + \mathcal{O}(k^{3/2}).\label{eq:vksol}
\end{equation}
The term of order $\mathcal{O}(k^{3/2})$ has a time dependence given by $a(\eta)\propto\eta^2$, and corresponds to the term proportional to $\alpha_k$ in (\ref{approxir}). 

Eventually, we have to match (\ref{eq:vksol}) with the explicit expression of  (\ref{approxir}) that one obtains with $a(\eta)=\qsubrm{\lambda}{0}\eta^{2}$ and $\eta_{\star}=-\infty$:
\begin{equation}
	\qsuprm{v}{IR}_k(\eta)=\alpha_k\qsubrm{\lambda}{0}\eta^2-\frac{\beta_k}{3\qsubrm{\lambda}{0}\eta}.\label{eq:vkint}
\end{equation}
By comparing \eqref{eq:vkint} with \eqref{eq:vksol}, one finds  
\begin{equation}
	\alpha_k=\mathcal{O}(k^{3/2})~~~\mathrm{and}~~~\beta_k=(3i/\sqrt{2})\qsubrm{\lambda}{0}k^{-3/2}.
\end{equation}
For infrared modes the contribution of $\alpha_k$ is negligible, so that \eqref{pir} simplifies to
\begin{equation}
\qsuprm{\qsubrm{\mathcal{P}}{T}(k)}{IR}=\frac{144 G}{\pi}\qsubrm{\lambda}{0}^2|I(\qsubrm{\eta}{e})|^2.\label{eq:pire}
\end{equation}
Therefore, in the IR limit we expect the power spectrum to be {\it scale invariant} (at least at the order of validity of our approximations). \\

The amplitude of the power spectrum in the IR regime is obtained by evaluating the integral $I(\qsubrm{\eta}{e})=\int^{\qsubrm{\eta}{e}}_{-\infty}{d\eta}/{a^2}$. In order to do this, we split the integral into three parts
\be
I(\qsubrm{\eta}{e})= I(-\infty,\qsubrm{\eta}{A})+I(\qsubrm{\eta}{A},\qsubrm{\eta}{i})+I(\qsubrm{\eta}{i},\qsubrm{\eta}{e}).\label{eq:split}
\ee
The first part  corresponds to the contracting phase, the second part corresponds to the bouncing phase, and the last part gives the contribution of the inflationary phase. With $a(\eta)=\qsubrm{\lambda}{0}\eta^2$ during the contracting phase, and recalling that  $\qsubrm{\eta}{A}=-[6/(m\qsubrm{\lambda}{0})]^{1/3}$, the first part of the integral is easy to compute:
\begin{equation}
	I(-\infty,\qsubrm{\eta}{A}) =\frac{m}{18\qsubrm{\lambda}{0}}.\label{eq:ia}
\end{equation}
The second part of the integral is first written in cosmic time, $I(\qsubrm{\eta}{A},\qsubrm{\eta}{i})=\int_{\qsubrm{t}{A}}^{\qsubrm{t}{i}}a(t)^{-3}dt$. During the bouncing phase we have found that $a=\qsubrm{a}{B}|y|^{-\tfrac{1}{3}}$. Then, with $y=\dot{x}/m$,  the integrand is proportional to $\dot{x}$ and the integral itself is proportional to the difference $|\qsubrm{x}{i}-\qsubrm{x}{A}|$ (which is given in (\ref{eq:xi})). Eventually, one gets
\be
	I(\qsubrm{\eta}{A},\qsubrm{\eta}{i})= -\frac{m}{18\qsubrm{\lambda}{0}}\frac{1}{\left|\cos\qsubrm{\theta}{A}\right|}\ln\left(\tfrac{1}{2}\Gamma\sqrt{\tfrac{\left|\cos\qsubrm{\theta}{A}\right|}{f}}\right).\label{eq:iai}
\ee

The last part of the integral corresponds to the slow-roll inflation as obtained from a massive scalar field. The calculations are well known in this case, leading to
\begin{equation}
	I(\qsubrm{\eta}{i},\qsubrm{\eta}{e})=\left(\frac{1}{3\qsubrm{a}{i}^3\qsubrm{H}{i}}-\frac{1}{3\qsubrm{a}{e}^3\qsubrm{H}{e}}\right)[1+\mathcal O(\qsubrm{\epsilon}{H})], \label{eq:iif}
\end{equation}
where $\qsubrm{\epsilon}{H}\equiv-\dot H/H^2$ is the slow-roll parameter which remains small in comparison to unity (except in the neighbourhood of $\qsubrm{t}{e}$). It will be neglected in the forthcoming calculations.  During slow-roll inflation, the Hubble parameter decreases linearly with cosmic time  while the scale factor grows exponentially. The second term, $1/(\qsubrm{a}{e}^3\qsubrm{H}{e})$, can be safely neglected as it is suppressed by a factor $\sim\exp(-3\qsubrm{N}{e})$, where  $\qsubrm{N}{e}$ denotes the number of e-folds from $\qsubrm{\eta}{i}$ to $\qsubrm{\eta}{e}$. This also means that the detailed dynamic of inflation is not needed here, since its contribution is rapidly negligible after a few e-folds. With the expressions of $\qsubrm{a}{i}$ and $\qsubrm{H}{i}$ given in \eqref{eq:hsr}, $I(\qsubrm{\eta}{i},\qsubrm{\eta}{e})$ evaluates to
\be
I\left(\qsubrm{\eta}{i},\qsubrm{\eta}{e}\right) =  \frac{m}{12\sqrt{3}{\qsubrm{\lambda}{0}}}\frac{\Gamma}{\left|\qsubrm{x}{i}\cos\qsubrm{\theta}{A}\right|}, \label{eq:iife}
\ee
with $x_i$ given in \eqref{eq:xi}. \\

Gathering the results \eqref{eq:ia}, \eqref{eq:iai} and \eqref{eq:iife}, the integral $I(\qsubrm{\eta}{e})$ can be written as $I\left(\eta_{\mathrm{e}}\right)=\frac{m}{18\qsubrm{\lambda}{0}}\left(1+\mathcal{I}+\mathcal{J}\right)$,
so that  the IR limit of the power spectrum \eqref{eq:pire} reads
\begin{equation}
\qsuprm{\qsubrm{\mathcal{P}}{T}(k)}{IR}=\frac{4 G}{9\pi}m^2|1+\mathcal{I}+\mathcal{J}|^2,\label{eq:pirdm}
\end{equation}
where
\begin{eqnarray}
\mathcal{I} & \equiv & -\frac{1}{\left|\cos\qsubrm{\theta}{A}\right|}\ln\left(\tfrac{1}{2}\Gamma\sqrt{\tfrac{\left|\cos\qsubrm{\theta}{A}\right|}{f}}\right),\label{eq:ical}\\
\mathcal{J} & \equiv & \frac{\Gamma\sqrt{3}}{2\left|\qsubrm{x}{i}\cos\qsubrm{\theta}{A}\right|}.\label{eq:jcal}
\end{eqnarray}
In general, $\mathcal{J}$ is much smaller than $\mathcal{I}$, suggesting that the contribution to the amplitude of the spectrum in the IR that corresponds to inflation is negligible (for instance with $\cos\qsubrm{\theta}{A}=1$ and $\Gamma=2\times10^{-7}$, one gets  ${\mathcal{J}}/{\mathcal{I}}\simeq0.002$). \\

The scale-invariance of the IR limit of the spectrum is a direct consequence of the fact that the infrared modes crossed the effective potential, $a''/a$, during the contracting phase whose dynamics is equivalent to that of a dust-like matter dominated era. No further assumption on the detailed dynamics of the bounce is needed to get the scale invariance (though the detailed dynamics is needed to get the amplitude of the power spectrum). The amplitude only depends on three parameters: the critical energy density, the mass of the scalar field, and the phase $\qsubrm{\theta}{A}$, between $x=m\phi/\sqrt{2\qsubrm{\rho}{c}}$ and $y=\dot{\phi}/\sqrt{2\qsubrm{\rho}{c}}$  at the start of the bouncing phase. The first two parameters are fundamental. The phase $\qsubrm{\theta}{A}$ depends on $\qsubrm{\theta}{0}$ which is a contingent parameter whose value sets the initial conditions (see \cite{linsefors1} for a more detailed discussion). The case $\cos\qsubrm{\theta}{A}\ll1$ may appear  problematic (as it would lead to a divergent power spectrum), however in this case the dynamic of the background would be different (with {\it deflation} before the bounce) and our analytical results would not be valid.

\subsection{Calculation of the UV limit}
\subsubsection{Definition of the UV regime}
By definition, the ultraviolet modes have remained well inside the Hubble radius until the phase of slow-roll inflation. They are insensitive to the background curvature during the contracting and the bouncing phase. The effective potential $a''/a$ can be written in terms of the Hubble parameter and its time derivative as ${a^{\prime\prime}}/{a}  =  a^{2}\left(\dot{H}+2H^{2}\right)$.
During the bounce, this expression becomes
\be
\frac{a^{\prime\prime}}{a}  =  \frac{8\pi G\qsubrm{\rho}{c}}{3}\qsubrm{a}{B}^2 y^{\frac{4}{3}}\left(4y^{2}-1\right).\label{eq:appoveraamp}
\ee
It is clear in \eqref{eq:appoveraamp} that the effective potential reaches  its maximum at the bounce, when $y=1$. This feature sets a scale, $\qsubrm{k}{UV} \equiv  \mathrm{max}\sqrt{{a^{''}}/{a}}$, which evaluates to 
\be
\qsubrm{k}{UV} = \qsubrm{a}{B}\sqrt{8\pi G\qsubrm{\rho}{c}}.
\ee
All modes with a wavenumber larger than $\qsubrm{k}{UV}$ crossed the potential during slow-roll inflation.  The UV limit of the power spectrum is defined by the modes with a wavenumber $k\gg\qsubrm{k}{UV}$.
\subsubsection{Primordial power spectrum in the UV regime}
In the dressed metric approach, the calculation of the UV limit of the power spectrum is straightforward. As during the bouncing phase the mode functions do not feel the curvature of space-time, they are well approximated by
\begin{equation}
	\qsuprm{v}{UV}_{k}(\eta)=\frac{1}{\sqrt{2k}}e^{ik\eta}~~~\mathrm{for}~\eta<\qsubrm{\eta}{i}.
\end{equation}
Once the Universe enters inflation, the term $a''/a$ cannot be neglected anymore and behaves as $(2+3\qsubrm{\epsilon}{H})/\eta^2$. The mode functions are now given by a linear combination of the Hankel functions of order $3/2+\qsubrm{\epsilon}{H}$. At this stage, the derivation of the primordial spectrum is simple: we have to match the Minkowski vacuum (well defined within the Hubble radius for $k\gg\qsubrm{k}{UV}$) with the mode functions commonly used in slow-roll inflation. The power spectrum in the UV regime is then given by the standard red-tilted power spectrum of slow-roll inflation (see \cite{lyth_stewart_1993,stewart_gong_2001,martin_schwarz_2000}),
\bse
\begin{eqnarray}
	\qsuprm{\qsubrm{\mathcal{P}}{T}(k)}{UV}&=&\frac{16G}{\pi}H^2 \left[1-2\qsubrm{\epsilon}{H}\left(2C+1\right)\right],\\
	\frac{\mathrm{d}\ln\qsuprm{\qsubrm{\mathcal{P}}{T}(k)}{UV}}{\mathrm{d}\ln k}&=&-2\qsubrm{\epsilon}{H},
\end{eqnarray}
\ese
where $H$ is the Hubble parameter evaluated when $k=aH$, and $C\simeq-0.73$.  At the order of validity of our approximation ($\Gamma\ll1$) and neglecting $\qsubrm{\epsilon}{H}$ in the amplitude, these expressions become
\bse
\bea
\qsuprm{\qsubrm{\mathcal{P}}{T}(k)}{UV}&=&\frac{16G}{\pi}m^2\left|\frac{\qsubrm{x}{i}}{\Gamma}\right|^2, \label{eq:puvdm}\\
\frac{\mathrm{d}\ln\qsuprm{\qsubrm{\mathcal{P}}{T}(k)}{UV}}{\mathrm{d}\ln k}&=&-6\left|\frac{\Gamma}{\qsubrm{x}{i}}\right|^2,\label{eq:si}
\eea
\ese
where $\Gamma\equiv m/\sqrt{24\pi G\qsubrm{\rho}{c}}$ and $\qsubrm{x}{i}$ is given by \eqref{eq:xi}. This prediction of a slightly red tilted spectrum  matches the standard inflationary model. The amplitude scales with $m^2$ and depends on the critical energy density in a non-trivial way. With $\cos\qsubrm{\theta}{A}=1$ for simplicity, one gets $\qsuprm{\qsubrm{\mathcal{P}}{T}(k)}{UV}\propto m^2\ln^2(m/\sqrt{G\qsubrm{\rho}{c}})$. Moreover, with the standard value $\Gamma=2\times10^{-7}$ we find that the spectral index at the start of inflation, given by \eqref{eq:si}, is $\qsubrm{n}{T}-1\simeq-0.007$.

\section{Power spectrum in the deformed algebra approach}
\label{sec:irda}
\subsection{The deformed algebra approach}
The calculations presented above can be extended to the case of the tensor power spectrum in the deformed algebra approach \cite{bojo1,thomCQG1,thomCQG2,Cailleteau:2012fy,barrau_bojo}. The equation of motion for the mode functions of the Mukhanov-Sasaki variables also takes the form of a Schr\"odinger equation. However,  the frequency term is time-dependent and the effective potential is different:
\begin{equation}
		v''_k(\eta)+\left(\Omega k^2-\frac{\qsubrm{z}{T}''}{\qsubrm{z}{T}}\right)v_k(\eta)=0, \label{eom2}
\end{equation}
where 
\be
\Omega\equiv1-2\frac{\rho}{\qsubrm{\rho}{c}} \,\,\,\mathrm{and}\,\,\,\,\qsubrm{z}{T}\equiv \frac{a}{\sqrt{\Omega}}. 
\ee
The region with $\Omega>0$, corresponding to $\rho<\qsubrm{\rho}{c}/2$, is Lorentzian whereas the region with $\Omega<0$, corresponding to $\rho>\qsubrm{\rho}{c}/2$, is Euclidean. Here, the mode functions are related to the amplitude of the tensor modes of the metric perturbation, $h_k$, via $v_k=\qsubrm{z}{T} h_k/\sqrt{32\pi G}$, so that the power spectrum is now defined as
\begin{equation}
\qsubrm{\mathcal{P}}{T}(k)=\frac{32 G k^3}{\pi}\left|\frac{v_k(\qsubrm{\eta}{e})}{\qsubrm{z}{T}(\qsubrm{\eta}{e})}\right|^2,
\end{equation}
where $\qsubrm{\eta}{e}$ denotes the conformal time at the end of slow-roll inflation. Actually, this definition is equivalent to (\ref{pdef}) because during slow-inflation $\qsubrm{z}{T}\simeq a$ (as a consequence of $\qsubrm{\rho}{i}\ll\qsubrm{\rho}{c}$).

\subsection{Calculation of the IR limit}
The IR limit  is defined exactly in the same way as in the dressed metric approach. From the expression of $a$ and $\rho$ as functions of  conformal time, one easily notices that $\Omega k^2-z''_\mathrm{T}/z_\mathrm{T}\simeq k^2-2/\eta^2+\mathcal{O}(\Gamma^2/\eta^{8})$. In the contracting phase, there is therefore no noticeable difference between the deformed algebra and the dressed metric approaches. The IR limit  still corresponds to modes with $k\ll \qsubrm{k}{IR}$, where $\qsubrm{k}{IR}$ is given in (\ref{eq:ka}).

The calculation of the IR limit of the spectrum proceeds in the same way as for the dressed metric approach. We first write the approximate solution to \eqref{eom2} 
in the infrared regime, 
\begin{equation}
	\qsuprm{v}{IR}_k(\eta)=\alpha_k\qsubrm{z}{T}(\eta)+\beta_k\qsubrm{z}{T}(\eta)\displaystyle\int^\eta_{-\infty}\frac{d\eta'}{\qsubrm{z}{T}^2(\eta')}+\mathcal{O}(k^2),\label{approxir2}
\end{equation}
from which the general expression of the IR limit of the power spectrum directly follows:
\begin{equation}
	\qsuprm{\qsubrm{\mathcal{P}}{T}(k)}{IR}=\frac{32 G k^3}{\pi}\left|\alpha_k+\beta_k\displaystyle\int^{\qsubrm{\eta}{e}}_{-\infty}\frac{d\eta}{\qsubrm{z}{T}^2}\right|^2.\label{eq:pomir}
\end{equation}
With the definition of $\qsubrm{z}{T}$, the integral on the RHS is simply given by the sum of $I(\qsubrm{\eta}{e})+I_\Omega(\qsubrm{\eta}{e})$, with $I(\qsubrm{\eta}{e})$ defined in (\ref{eq:int}), and
\be
I_\Omega(\qsubrm{\eta}{e})\equiv-2\displaystyle\int^{\qsubrm{\eta}{e}}_{-\infty}\frac{\rho}{\qsubrm{\rho}{c}}\frac{d\eta}{a^2}.
\ee
As before, the two constants $\alpha_k$ and $\beta_k$  in \eqref{eq:pomir} are obtained by matching the solution (\ref{approxir2}) with a set of solutions valid for any wavenumber $k$ during the contracting phase. During the contracting phase the difference between  $(\Omega k^2-\qsubrm{z}{T}^{\prime\prime}/\qsubrm{z}{T})$ and  $(k^2-a''/a)$ can be neglected. Consequently, the two constants $\alpha_k$ and $\beta_k$ take the same value as before: $\alpha_k=\mathcal{O}(k^{3/2})$ and $\beta_k=(3i/\sqrt2)\qsubrm{\lambda}{0}k^{-3/2}$. With these expressions, the IR limit of  the power spectrum becomes
\begin{equation}
\qsuprm{\qsubrm{\mathcal{P}}{T}(k)}{IR}=\frac{144 G }{\pi}\qsubrm{\lambda}{0}^2\left|I(\qsubrm{\eta}{e})+I_\Omega(\qsubrm{\eta}{e})\right|^2.
\end{equation}
Note that all the differences between the deformed algebra and the dressed metric approach are encoded in the integral $I_\Omega(\eta_\mathrm{e})$.

Now, we will show that $I_\Omega(\qsubrm{\eta}{e})/I(\qsubrm{\eta}{e})=\mathcal O (\Gamma^2)$, so that the contribution of $I_\Omega(\qsubrm{\eta}{e})$  to the IR limit of the spectrum can be neglected. First, we split the integral into three parts: $I_\Omega(\qsubrm{\eta}{e})=I_{\Omega}(-\infty,\qsubrm{\eta}{A})+I_\Omega(\qsubrm{\eta}{A},\qsubrm{\eta}{i})+I_\Omega(\qsubrm{\eta}{i},\qsubrm{\eta}{e})$. The proof is straightforward for the first and the third parts of the integral. Indeed, recalling that before $\qsubrm{\eta}{A}$ the energy density remains smaller than $\qsubrm{\rho}{A}=\Gamma^2\qsubrm{\rho}{c}$, we have
\be
	I_\Omega(-\infty,\qsubrm{\eta}{A})\equiv-2\displaystyle\int^{\qsubrm{\eta}{A}}_{-\infty}\frac{\rho}{\qsubrm{\rho}{c}}\frac{d\eta}{a^2}\leq-2\Gamma^2I(-\infty,\qsubrm{\eta}{A}). \label{eq:ioa}
\ee
The same holds for $I_\Omega(\qsubrm{\eta}{i},\qsubrm{\eta}{e})$, with $\qsubrm{\rho}{i}$ instead of $\qsubrm{\rho}{A}$. The remaining part  of the integral is 
\begin{equation}
	I_\Omega(\qsubrm{\eta}{A},\qsubrm{\eta}{i})\equiv-2\displaystyle\int^{\qsubrm{\eta}{i}}_{\qsubrm{\eta}{A}}\frac{\rho}{\qsubrm{\rho}{c}}\frac{d\eta}{a^2}.
\end{equation}
During the bouncing phase, $\rho=\qsubrm{\rho}{c}y^2$ and $a=\qsubrm{a}{B}|y|^{-\frac{1}{3}}$ so when we switch to cosmic time, the integral becomes 
\begin{equation}
I_\Omega(\qsubrm{\eta}{A},\qsubrm{\eta}{i})=-(2/\qsubrm{a}{B}^2)\displaystyle\int^{\qsubrm{t}{i}}_{\qsubrm{t}{A}}|y(t)|^3dt.
\end{equation}
The last step is to express $dt$ in terms of $dy$ with \eqref{eq:sfa}. Then the integration can be performed analytically and leads to
\be
I_\Omega(\qsubrm{\eta}{A},\qsubrm{\eta}{i})=-\frac{m}{36\qsubrm{\lambda}{0}}\frac{\sin^2{\qsubrm{\theta}{A}}}{\cos{\qsubrm{\theta}{A}}}\Gamma^2.
\ee
Therefore, at order $\mathcal{O}(\Gamma^2)$, we predict no difference for the IR limits of the power spectra in both approaches. The IR limit of the power spectrum in the deformed algebra approach is still given by \eqref{eq:pirdm}.

\subsection{Calculation of the UV limit}
The UV limit of the power spectrum in the deformed algebra approach has already been discussed in \cite{linsefors2}, here we recall the conclusion of this previous work. Thanks to numerical integrations for the equation of motion as well as WKB based arguments, it is shown that the primordial power spectrum exponentially grows with the wavenumber $k$, for large values of $k$. Actually, oscillations are still superimposed to this exponential envelope. During the bouncing phase, the term $\qsubrm{z}{T}^{\prime\prime}/\qsubrm{z}{T}$ reaches a maximum $|\qsubrm{z}{T}^{\prime\prime}/\qsubrm{z}{T}|_{t_B}=40\pi G\qsubrm{\rho}{c}$. This means that for modes such that $k^2\gg40\pi G\qsubrm{\rho}{c}$, the time-dependent frequency in the equation of motion, $\Omega k^2-\qsubrm{z}{T}^{\prime\prime}/\qsubrm{z}{T}$, is dominated by $\Omega k^2$ during most of the cosmic history prior to inflation. The Euclidean phase around the bounce, $\Omega<0$, leads to an instability in the equation of motion so the amplitude of tensor modes recieves a real exponential contribution, {\it i.e.}~$v_{k\to\infty}\propto \exp({k\times\int_{\Delta\eta}\sqrt{\left|\Omega\right|}d\eta})$, where the integration is performed over the interval $\Delta\eta$ corresponding to the Euclidean phase.

 %%%%%%%%%%%%%%%%%%%%%%%%%%%

\section{Power spectrum at all scales: numerical results}
\label{sec:numpow}
Deriving the primordial power spectrum for tensor modes at all scales requires a numerical integration of {\it both} the equation of motion of the mode functions ((\ref{eom}) or (\ref{eom2}) depending on the approach) {\it and} the equations of motion for the background (gathered in the system of equations (\ref{eq:sys})). The numerical integration is performed starting in the contracting phase, when ${\qsubrm{\rho}{0}}/{\qsubrm{\rho}{c}}\ll\Gamma^2$. For the background, the initial conditions are set by choosing the value of $\qsubrm{\theta}{0}$,  the initial phase between the share of potential energy and kinetic energy in the total energy density. 
For the perturbations, the initial conditions are set during the contracting phase when the modes are well inside the horizon. The initial state of the perturbation can then be identified with the usual Minkowski vacuum.  \\

The detailed dynamics of the background ({\it e.g.} the value of $x$ at the bounce, or the number of e-folds during inflation) and subsequently the detailed shape of $\qsubrm{\mathcal{P}}{T}(k)$, are fully determined by two types of parameters: the mass of the scalar field and the critical energy density on one hand, and the phase $\qsubrm{\theta}{0}$ on the other hand.  

The mass of the scalar field and the critical energy density can be seen as fundamental constants of the model. Though their values are not  known, some particular values are favored by CMB observations and theoretical considerations.  Even if the details of the calculation using the minimal area gap of LQG still need clarification, some  dimensional arguments lead to believe that the value of $\qsubrm{\rho}{c}$ should not be far from the Planck scale. Note that  $\qsubrm{\rho}{c}$ is the only parameter linked to LQG (via its dependence on the Immirzi parameter, $\gamma$). The value commonly accepted is $\qsubrm{\rho}{c}=0.41\qsubrm{m}{Pl}^4$, and we shall use it as the standard choice in our numerical simulation. The value of the mass of the scalar field, as deduced from the CMB observations, is generally chosen to be $m\simeq1.2\times10^{-6} \qsubrm{m}{Pl}$ \cite{liddle}.

The parameter $\qsubrm{\theta}{0}$ has a different status since it is totally {\it contingent}. Its value can vary between $0$ and $2\pi$ (actually the range $0<\qsubrm{\theta}{0}<\pi$ is enough as the equations remain unchanged under the transformation $\qsubrm{\theta}{0}\to\qsubrm{\theta}{0}+\pi$). As underlined in \cite{linsefors1}, most of the values of $\qsubrm{\theta}{0}$ lead to a universe with a phase of inflation shortly after the bounce (and no deflation before the bounce). We have restricted ourselves to this kind of solutions since they are the most probable, and the more in line with our current knowledge of the cosmic history (believed to have underwent a phase of primordial inflation).\\

Qualitatively, we can already anticipate the global shape of the primordial power spectrum. Irrespective of any approach, its shape is driven by the background evolution through the functions $a$ and $\Omega$ (in the deformed algebra approach) and their time derivatives. Our analysis is restricted to the wide range of cosmic histories that split into three main eras: a classical (dust-like) contracting phase, a bouncing phase where quantum effects are significant, and  a classical inflationary phase. We anticipate the shape of the primordial power spectrum to be {\it qualitatively} unaffected by the values of $m$, $\qsubrm{\rho}{c}$, and $\qsubrm{\theta}{0}$. (Obviously, the precise values of these parameters will affect the scales and amplitudes involved in the spectrum at a {\it quantitative} level). We can also anticipate three regimes in the power spectra, corresponding to: the modes that have left the horizon during the contracting phase (large scales); the modes that have left the horizon during the bouncing phase (intermediate scales); the modes that have remained within the horizon until the start of the inflationary phase (small scales). For the large and small scales, we should recover the IR and UV limit derived analytically in the previous sections. \\

In the next three sections we present the primordial power spectra obtained within each approach. We study the influence of the three parameters, $m$, $\rho_c$, and $\theta_0$. For each varying parameter, we present the primordial power spectra as predicted by each approach, thus facilitating the comparison.

We use Planck units hereafter, with the following definition of the Planck mass, $\qsubrm{m}{Pl}=1/\sqrt{G}$. For simplicity, we normalize the scale factor at the time of the bounce, setting $\qsubrm{a}{B}=1$. The power spectra are depicted as functions of the comoving wavenumber $k$.

\subsection{Varying the mass of the scalar field}
\label{ssec:mass}
\begin{figure}
\begin{center}
	\includegraphics[scale=0.4]{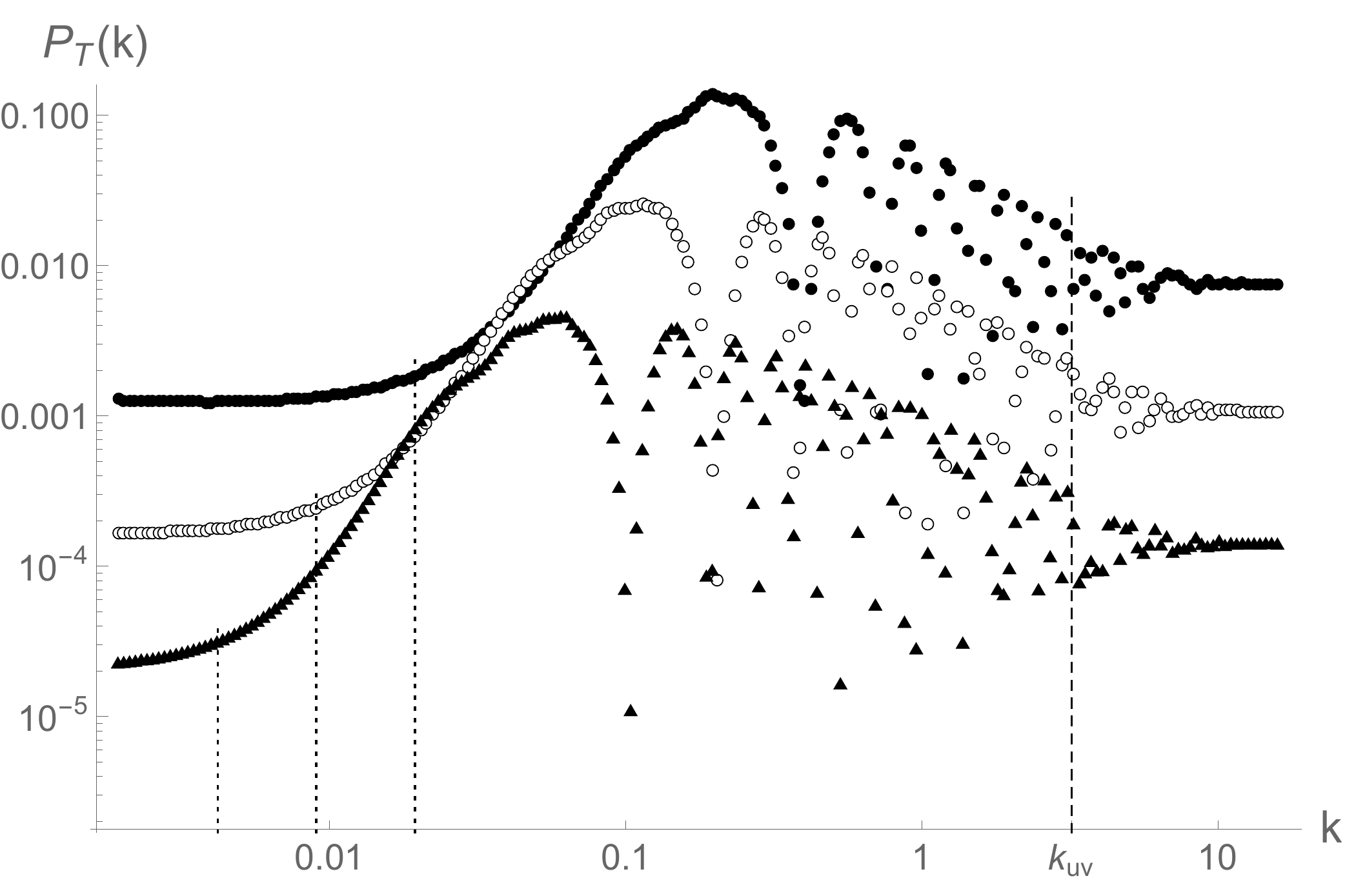}\\
	\includegraphics[scale=0.4]{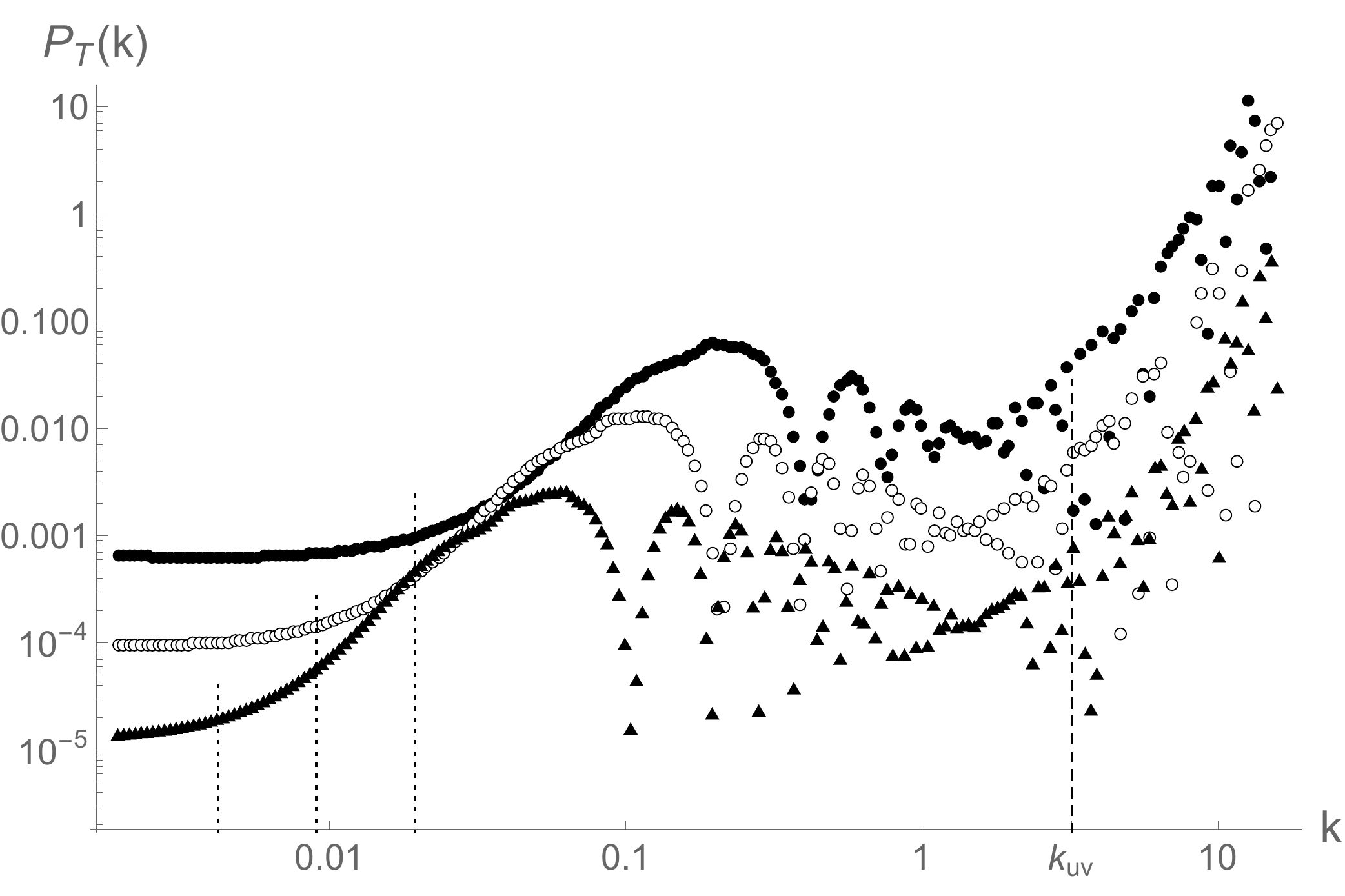}
\caption{Primordial power spectra for  tensor modes in the dressed metric approach (upper panel) and in the deformed algebra approach (lower panel) for different values of the mass of the scalar field. The critical energy density is $\qsubrm{\rho}{c}=0.41\qsubrm{m}{Pl}^4$ and $\cos\qsubrm{\theta}{A}\simeq1$. The mass of the scalar field takes three values: $m=10^{-3}\qsubrm{m}{Pl}$ (triangles), $m=10^{-2.5}\qsubrm{m}{Pl}$ (open disks), and $m=10^{-2}\qsubrm{m}{Pl}$ (black disks). The dashed vertical line at large $k$, corresponds to $\qsubrm{k}{UV}$ (which does not depend on $m$). The dotted vertical lines at smaller $k$ correspond to $\qsubrm{k}{IR}$ (which scales as $m^{2/3}$).}
\label{fig:mass}
\end{center}
\end{figure}

The primordial power spectrum for different values of the mass of the scalar field is presented in Fig.~\ref{fig:mass}. The upper panel corresponds to the dressed metric approach and the lower panel to the deformed algebra approach. The mass takes three different values: $m=10^{-3}\qsubrm{m}{Pl}$ (triangles), $m=10^{-2.5}\qsubrm{m}{Pl}$ (open disks), and $m=10^{-2}\qsubrm{m}{Pl}$ (black disks).  For numerical convenience, these values are larger than the preferred value. However, the results can be extrapolated and the associated phenomenology shall be studied with values closer to $10^{-6}\qsubrm{m}{Pl}$ \cite{boris}.  The critical energy density is set equal to $0.41\qsubrm{m}{Pl}^4$ and $\cos\qsubrm{\theta}{A}\simeq1$.\\

In the dressed metric approach (upper panel of Fig.~\ref{fig:mass}), there are three regimes in the primordial power spectrum:
\begin{enumerate}[(i)]
\item At the largest scales, for $k<\qsubrm{k}{IR}$ with
$$
\qsubrm{k}{IR}=\frac{1}{3\sqrt{2}}\left(\frac{m^2\sqrt{24\pi G\qsubrm{\rho}{c}}}{\left|\cos\qsubrm{\theta}{A}\right|}\right)^{1/3},
$$
the power spectrum is scale invariant in agreement with the analytical calculations of Sec.~\ref{sec:irpow}. This corresponds to modes that were amplified mainly during the classical contracting phase. In Fig.~\ref{fig:mass}, the scale corresponding to $\qsubrm{k}{IR}$ is depicted with vertical dotted lines. At this scale, there is a transition in the numerical results that  is in perfect agreement with the analytical formula (especially its $m^{2/3}$ dependence). Moreover, it is clear on the figure that the numerical IR limit of the spectrum behaves as $m^2$, again in perfect agreement with (\ref{eq:pirdm}). 
\item For intermediate scales, such that $\qsubrm{k}{IR}<k<\qsubrm{k}{UV}$, with 
$$
\qsubrm{k}{UV}=\sqrt{8\pi G \qsubrm{\rho}{c}},
$$
the amplitude of the power spectrum is oscillating. This part of the spectrum corresponds to modes that were amplified during the bouncing phase. The first peak corresponds to a maximum of the power spectrum, that reaches  $100\times\qsuprm{\qsubrm{\mathcal{P}}{T}}{IR}$ approximately. Then, the amplitude of the oscillations is damped for increasing values of the wavenumber. Intuitively, these oscillations can be understood as due to quasi-bound states in the effective Schr\"odinger equation. The second transition scale, $\qsubrm{k}{UV}$, is depicted in Fig.~\ref{fig:mass} as a vertical dashed line and is in agreement with the transition scale found numerically. 
\item At the smallest scales, $k>\qsubrm{k}{UV}$, the power spectrum is a power law with a slightly red spectral index, just as predicted by the standard inflationary paradigm. This part of the power spectrum corresponds to modes that have remained inside the horizon until the start of the inflationary phase. The numerical results are in agreement with the UV limit derived analytically  in Sec.~\ref{sec:irpow}, see Fig.~\ref{fig:theta}. \\
\end{enumerate}

In the deformed algebra approach the primordial tensor power spectrum also features three different regimes (see the lower panel of Fig.~\ref{fig:mass}). The first two regions ({\it i.e.}~the large scales, $k<\qsubrm{k}{IR}$, and the intermediate scales, $\qsubrm{k}{IR}<k<\qsubrm{k}{UV}$), are almost identical to the power spectrum derived in the dressed metric approach. The scale-dependence of the power spectrum and the transition scales are the same. This is because the impact of $\Omega$ is subdominant for these modes. However, within these two regions the numerical results suggest that for $k<\qsubrm{k}{UV}$ the amplitude of the spectrum in the deformed algebra approach is slightly smaller than in the dressed metric approach (by less than a factor 2). (This feature could not be explained by our analytics.)

At smaller scales, $k>\qsubrm{k}{UV}$, the primordial power spectrum in the deformed algebra approach strongly differs from the one predicted by the dressed metric approach. As already suggested by our analytical considerations (see Sec.~\ref{sec:irda}), the power spectrum is exponentially increasing with the wavenumber (as a result of the instability generated by $\Omega$ which is negative-valued around the bounce), with superimposed oscillations.
Note that the numerical results confirm once again that the scale defining the transition between the intermediate scales (oscillations) and the large scales (exponential growth) does not depend on $m$. 

The UV behavior of the spectrum clearly raises questions. The first one is related to the fundamentally trans-Planckian nature of these modes. As demonstrated in \cite{agullo1}, this is not a problem when considering the appropriate length operator in loop quantum gravity. A more serious issue is related to the use of the perturbation theory when the spectrum increases exponentially. Obviously, backreaction should be taken into account in this regime and the results shown here are not fully reliable anymore. They just give a general trend and not the accurate shape of the spectrum. However, we believe that this is basically enough for the phenomenological purposes we are interested in. The most interesting region, that is the oscillatory one, is under control and the $C_{\ell}$ CMB spectrum can be safely calculated  \cite{boris}. If the observational window of wavenumber was to fall on the exponentially rising part, this would anyway lead to a situation incompatible with data (as the tensor to scalar ratio is small). The perturbation theory breaks at a level where tensor modes are anyway excluded by current data.

\subsection{Varying the critical energy density}
\label{ssec:rho}
The critical energy density depends on the Immirzi parameter, a fundamental parameter in LQG, whose value is traditionally deduced from a calculation of the black holes entropy. Nonetheless, it  has been recently argued \cite{bianchi} that the formula for the entropy of black holes can be recovered, in the framework of LQG, without specifying the value of the Immirzi parameter. Recently, a quasi-local description of a black hole \cite{Frodden:2011eb} was indeed shown to allow one to recover at the semi-classical limit the expected thermodynamical behaviors of a black hole for all values of $\gamma$ \cite{Ghosh:2011fc}, assuming the existence of a non trivial chemical potential conjugate to the number of horizon punctures. A detailed microscopic mechanism was also put forward in \cite{Frodden:2012dq} and \cite{analytic} where the area degeneracy was analytically continued from real $\gamma$ to complex $\gamma$  and evaluated at the complex values $\gamma=\pm i$. This motivates us to consider other values for the critical energy density and discuss how it can affect the primordial tensor power spectrum. \\

\begin{figure}
\begin{center}
	\includegraphics[scale=0.4]{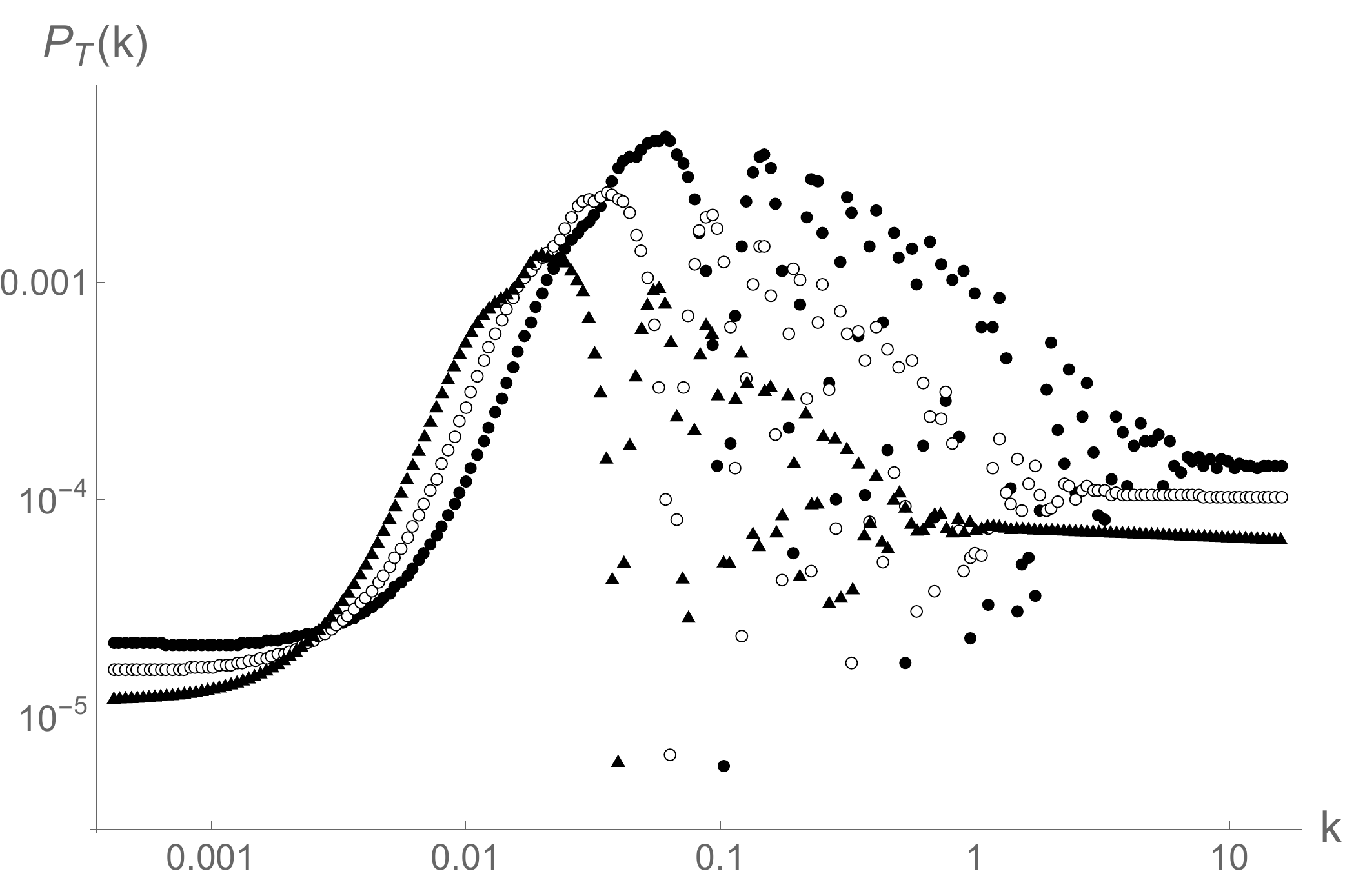}\\
	\includegraphics[scale=0.4]{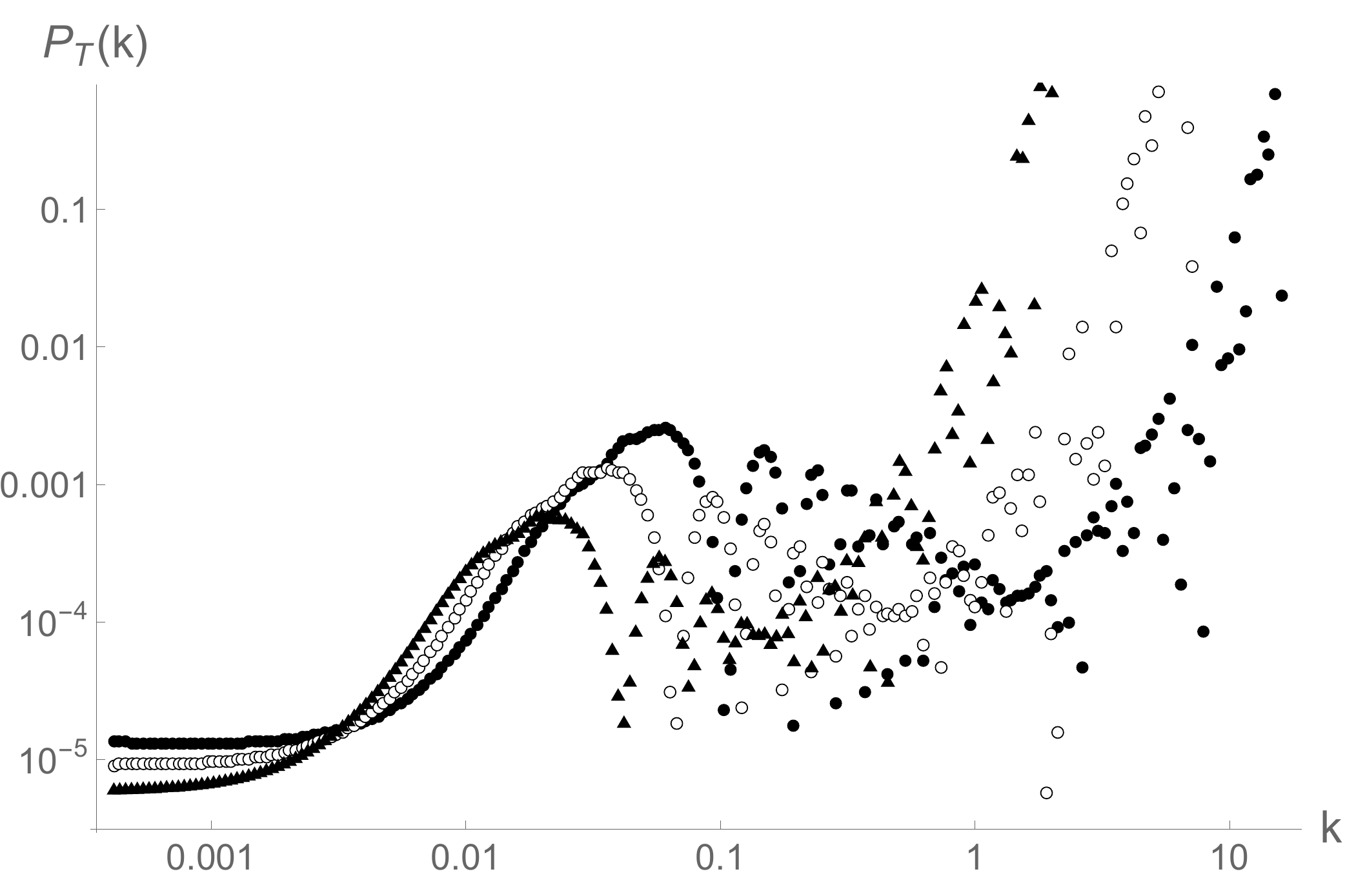}
\caption{Primordial power spectra for the tensor modes in the dressed metric approach (upper panel) and in the deformed algebra approach (lower panel) for different values of $\qsubrm{\rho}{c}$. The mass of the scalar field is $m=10^{-3}\qsubrm{m}{Pl}$ and $\cos\qsubrm{\theta}{A}\simeq1$. The critical energy density is $\qsubrm{\rho}{c}=0.0041\qsubrm{m}{Pl}^4$ (triangles), $\qsubrm{\rho}{c}=0.041\qsubrm{m}{Pl}^4$ (open disks) and $\qsubrm{\rho}{c}=0.41\qsubrm{m}{Pl}^4$ (black disks).}
\label{fig:rho}
\end{center}
\end{figure}

In Fig.~\ref{fig:rho}, we show the primordial tensor power spectra for different values of $\qsubrm{\rho}{c}$. Here, the mass of the scalar field is set equal to $m=10^{-3}\qsubrm{m}{Pl}$, and $\cos\qsubrm{\theta}{A}\simeq1$. The upper panel corresponds to the dressed metric approach and the lower panel to the deformed algebra approach. The different values of the critical energy density are $\qsubrm{\rho}{c}=0.0041\qsubrm{m}{Pl}^4$ (triangles), $\qsubrm{\rho}{c}=0.041\qsubrm{m}{Pl}^4$ (open disks), and $\qsubrm{\rho}{c}=0.41\qsubrm{m}{Pl}^4$ (black disks) which is the theoretically favored value. 

The global shape of the primordial power spectrum is recovered for both approaches, with three different regions. The positions of the transition scales, $\qsubrm{k}{IR}$ and $\qsubrm{k}{UV}$, clearly depend on $\qsubrm{\rho}{c}$ irrespectively of the approach. The IR transition scale, $\qsubrm{k}{IR}$, mildly decreases for smaller values of $\qsubrm{\rho}{c}$, in agreement with the analytical calculations that led to $\qsubrm{k}{IR}\propto (G\qsubrm{\rho}{c})^{1/6}$. The UV transition scale, $\qsubrm{k}{UV}$, is more strongly dependent on the value of the critical energy density, also in agreement with the scaling derived analytically, $\qsubrm{k}{UV}\propto\sqrt{G\qsubrm{\rho}{c}}$.

For the dressed metric approach, a decrease of $\qsubrm{\rho}{c}$ yields a slight decrease of the amplitude of the primordial power spectrum at {\it all} scales. This feature is also suggested by the analytical results, as both formulae for the UV and IR limits depend on the critical energy density  as $\sim\ln^2(m/\sqrt{G\qsubrm{\rho}{c}})$. \\

For the deformed algebra approach, a decrease of $\qsubrm{\rho}{c}$ leads to a slight decrease of the amplitude of the  spectrum at large and intermediate scales as in the dressed metric approach. At smaller scales, $k>\qsubrm{k}{UV}$, the smallest value of $\qsubrm{\rho}{c}$ corresponds to the fastest divergence of the spectrum. Analytically, we expect this divergence to scale as $\propto \exp({k\int_{\Delta\eta}\sqrt{\left|\Omega\right|}d\eta})$,  where the interval $\Delta\eta$ corresponds to the euclidean phase. Therefore we can define the rate of growth of the spectrum in the UV as $k_\Omega \equiv 1/\int_{\Delta\eta}\sqrt{\left|\Omega\right|}d\eta$. 

This integral can be computed, leading to 
\be
k_\Omega\simeq0.8\sqrt{24\pi G\qsubrm{\rho}{c}}\label{eq:kc}.
\ee
So, small values of the critical energy density indeed correspond to a quicker divergence of the power spectrum in the UV.

\subsection{Dependence on $\qsubrm{\theta}{0}$}
\label{ssec:theta}

\begin{figure}
\begin{center}
	\includegraphics[scale=0.4]{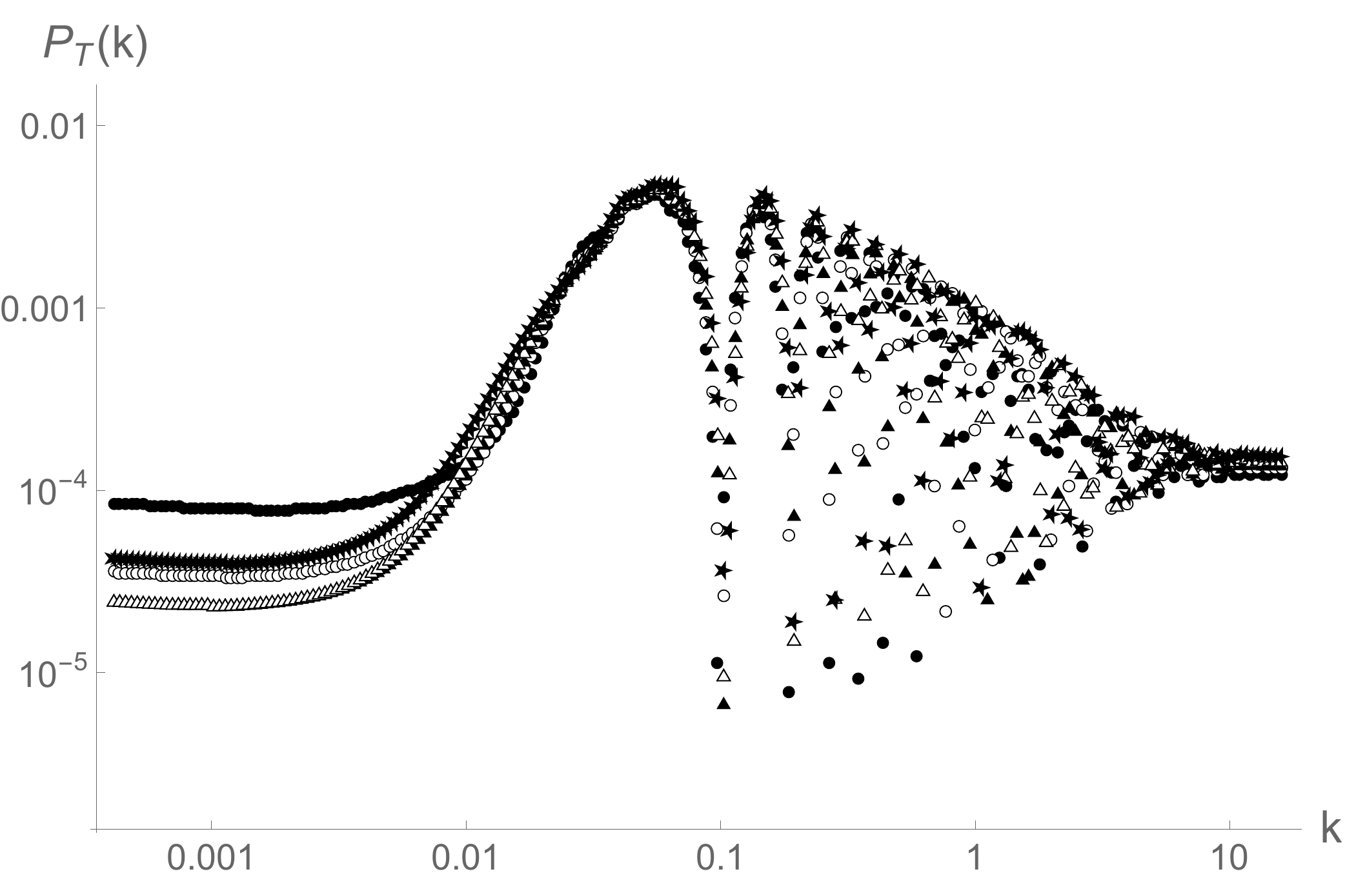}\\
	\includegraphics[scale=0.4]{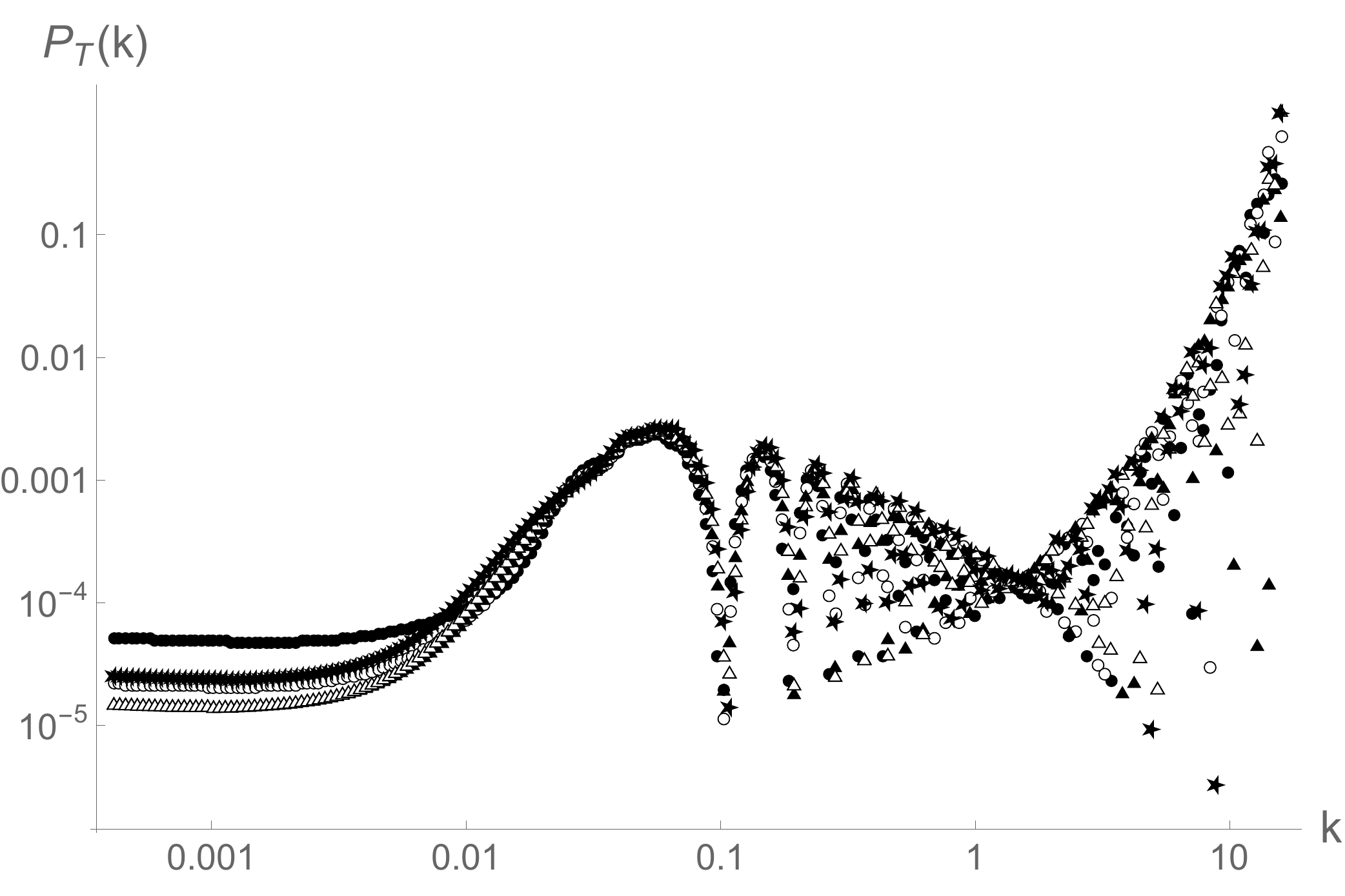}
	\caption{Primordial power spectra for the tensor modes in the dressed metric approach (upper panel), and in the deformed algebra approach (lower panel). The parameter $\qsubrm{\theta}{0}$ varries from $(\pi/2-1)\simeq0.18\times\pi$ to $(\pi/2+1)\simeq0.81\times\pi$. The exact values are: plain disks for $\qsubrm{\theta}{0}=(\pi/2-1)$, open disks for $\qsubrm{\theta}{0}=(\pi/2-1/2)$, plain triangles for $\qsubrm{\theta}{0}=\pi/2$, open triangles for $\qsubrm{\theta}{0}=(\pi/2+1/2)$, and plain stars for $\qsubrm{\theta}{0}=(\pi/2+1)$. The mass of the field is $m=10^{-3}\qsubrm{m}{Pl}$, and the critical energy density is $\qsubrm{\rho}{c}=0.41\qsubrm{m}{Pl}^4$.} 
	\label{fig:theta}
\end{center}
\end{figure}
The primordial power spectra for different choices of $\qsubrm{\theta}{0}$ are shown in Fig.~\ref{fig:theta},  in the dressed metric approach (upper panel) and in the deformed algebra approach (lower panel). The mass of the scalar field is $m=10^{-3}\qsubrm{m}{Pl}$, and the critical energy density is $\qsubrm{\rho}{c}=0.41\qsubrm{m}{Pl}^4$. We chose five values of $\qsubrm{\theta}{0}$,  equally spaced between $(\pi/2-1)$ and $(\pi/2+1)$, ensuring that the background goes through a phase of inflation after the bounce. In the numerical simulations we have always set $\alpha=17\pi/4+1$, so that $\cos\qsubrm{\theta}{A}=0$ (with $\qsubrm{\theta}{A}=2(\alpha-1)+\qsubrm{\theta}{0}$) corresponds to $\qsubrm{\theta}{0}=0$ and $\alpha$ is significantly larger than one.

The main impact of $\qsubrm{\theta}{0}$ is in the IR regime. Both the infrared transition scale $\qsubrm{k}{IR}$ (varying as $\sim\left|\cos(\qsubrm{\theta}{A})\right|^{-1/3}$) and the amplitude of the IR limit of the spectrum are significantly depending on $\qsubrm{\theta}{0}$. This is true in both approaches. At intermediate and smaller scales, the power spectra are nearly independent of the choice of $\qsubrm{\theta}{0}$, again irrespectively of the considered approach. The numerical results confirm that the ultraviolet transition scale $\qsubrm{k}{UV}$ is independent of $\qsubrm{\theta}{0}$. Moreover, in the deformed algebra approach the growth rate of the spectrum in the UV appears to be independent on  $\qsubrm{\theta}{0}$ too, in agreement with \eqref{eq:kc}. \\

\begin{figure}
\begin{center}
	\includegraphics[scale=0.4]{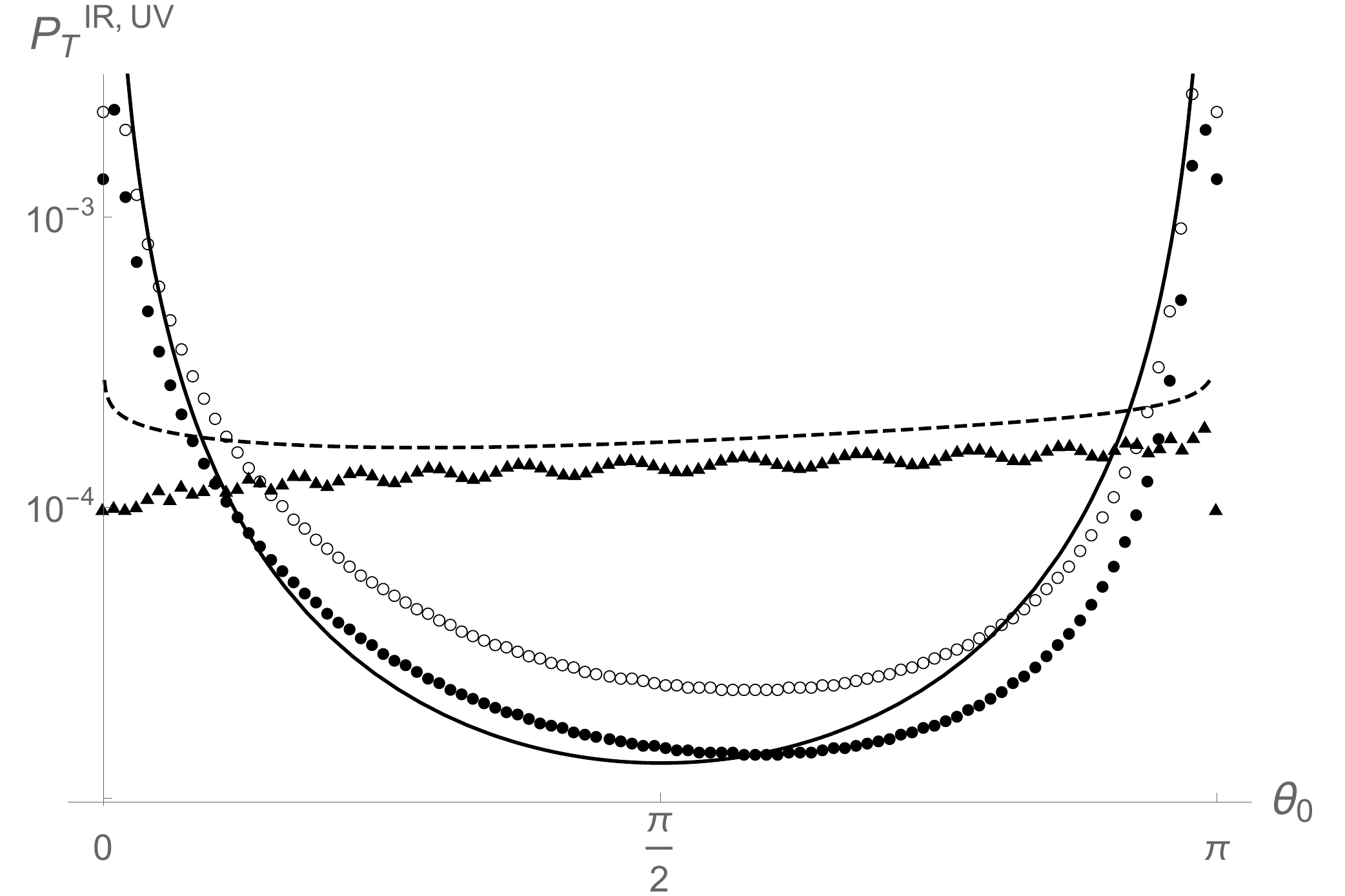}
	\caption{The infrared limit (disks) and the ultraviolet limit (triangles) of the primordial power spectrum as a function of $\qsubrm{\theta}{0}$. The solid black curve corresponds to the analytical calculations for $\qsuprm{\qsubrm{\mathcal{P}}{T}}{IR}$, see (\ref{eq:pirdm}). The IR limit from a numerical simulation is displayed with open disks for the dressed metric approach, and black disks in the deformed algebra approach. The dashed black curve stands for the analytical UV limit in the dressed metric approach, see (\ref{eq:puvdm}). The UV limit in the dressed metric approach, as derived from the numerics, corresponds to the black triangles. The mass of the scalar field is $m=10^{-3}\qsubrm{m}{Pl}$, and the critical energy density is $\qsubrm{\rho}{c}=0.41\qsubrm{m}{Pl}^4$.}
	\label{fig:iruv}
\end{center}
\end{figure}

In order to highlight the dependence of the IR limit as a function of $\theta_0$, Fig.~\ref{fig:iruv} shows this limit in both approaches and for different choices of $\qsubrm{\theta}{0}$, with $m=10^{-3}\qsubrm{m}{Pl}$, and $\qsubrm{\rho}{c}=0.41\qsubrm{m}{Pl}^4$. The solid black curve corresponds to the analytical calculation for $\qsuprm{\qsubrm{\mathcal{P}}{T}}{IR}$, see (\ref{eq:pirdm}). This analytical curve is valid for both the dressed metric and the deformed algebra approaches at first order in $\Gamma\equiv m/\sqrt{24\pi G\qsubrm{\rho}{c}}$. The numerical derivation of the IR limit is displayed as open disks for the dressed metric approach, and as black disks in the deformed algebra approach. We observe a fairly good agreement between analytical and numerical results. Although there are some differences in the amplitude of the IR limit\footnote{The discrepancy is not surprising. First of all, the analytic result is based on some approximations for the time dependence of $a$ and $\mathbf\Omega$. Second, the numerical evaluation fo $\qsuprm{\qsubrm{\mathcal{P}}{T}}{IR}$  cannot be exactly obtained for $k=0$ since this would require to start the numerical integration at $\eta\to-\infty$ which is unfeasible. We believe that these features are at the origin of the disagreement between the numerics and the analytics.}, the behavior as a function of $\qsubrm{\theta}{0}$ is consistent between the analytics and the numerics. This shows that $\qsuprm{\qsubrm{\mathcal{P}}{T}}{IR}$ strongly depends on $\qsubrm{\theta}{0}$, the former varying by more than one order of magnitude from its minimal value at $\qsubrm{\theta}{0}=\pi/2$ (thus giving $\cos\qsubrm{\theta}{A}=1$), to its maximal value reached when $\qsubrm{\theta}{0}$ tends to 0 or $\pi$.

In the restricted case of the dressed metric approach, the UV limit as a function of $\theta_0$  is also displayed in Fig.~\ref{fig:iruv}. The dashed black curve stands for the analytical calculation presented in (\ref{eq:puvdm}). The UV limit obtained from the numerical simulation is displayed with triangles. A good agreement is also observed here. Nonetheless, the remaining difference between the analytical  and numerical results certainly comes from the approximations involved in the determination of $\qsubrm{x}{i}$, on which the UV limit depend.

\section{Conclusion and discussion}
\label{sec:conc}

In this work, we have compared the dressed metric and deformed algebra approaches to loop quantum cosmology. In order to compare them efficiently, we have set the initial conditions in the same way for both approaches (in the remote past of the classical contracting branch). This is consistent and arguably the most obvious choice if the word \textit{initial} is to be taken literally. It is however fair to mention that this is not the only choice one could have made. As far as the dressed metric approach is concerned, the authors who developed the strategy have preferred to set the initial conditions at the bounce \cite{agullo1,agullo2,agullo3}. Then the initial state for tensor perturbations is given by a 4th-order WKB vacuum defined for $k\geq \qsubrm{k}{UV}$. In fact, their results seem to be very similar to ours (for the range of scales covered by both choices of initial conditions). As far as the deformed algebra approach is concerned, it should be underlined that it is also possible to set initial conditions at the surface of signature change. This has been investigated in \cite{jakub} and leads to a different spectrum. If these issues are left for future considerations and if we focus on the comparison with similar initial conditions, several important conclusions can already be drawn.

First, it is remarkable that for both approaches the IR limit is the same and basically agrees with the prediction of standard general relativity. Therefore at the largest scales, the primordial tensor power spectrum cannot be used to probe quantum gravity (at least in this setting).

Second, there is a strong difference between the approaches in the ultraviolet regime. Whereas the dressed metric simply leads to the slightly red-tilted power spectrum, as predicted in standard inflationary cosmology, the deformed algebra leads to an exponentially increasing spectrum (modulated by oscillations).

Third, at intermediate scales, a very interesting behavior appears. Not only because it is substantially different from the predictions of the standard inflationary models but also because both predictions are in agreement with each other! This region seems to exhibit a {\it universal LQC effect} that has been searched for during the last decade. In addition, the phase of the oscillations that appear at these intermediate scales, does {\it not} depend on the unknown (and fundamentally random) phase parameter, $\qsubrm{\theta}{0}$. This opens an interesting avenue in the perspective of testing the predictions of effective LQC.\\

In the future, this work should be extended in two directions. One is to consider not only the tensor modes, that have not yet been observed, but also the well known scalar modes. The relevant equations have already been derived for the dressed metric approach but are still to be investigated into more details in the deformed algebra approach. The reason for this difficulty is related to divergences (at the bounce and at the change of signature) that should be regularized. The difficulty is however more technical than conceptual and should be solved soon. 

The second path to follow is naturally to go more deeply into the phenomenology of this comparison and calculate the corresponding cosmic microwave background $C_\ell$ spectra which are already constrained by observations. Two main tasks will have to be pursued. The first is related to the number of e-folds that the Universe underwent since the bounce. This number depends, among other parameters, on $\qsubrm{\theta}{0}$ and on the reheating temperature. Once the number of e-folds since the bounce will be specified, the range of wavenumbers considered in this study that falls within our observable window will be completely determined. The second important task is to investigate how the oscillations at intermediate scales shall be washed out by the transfer phenomena that occur between the end of inflation and the decoupling. \\

\acknowledgments

BB is supported by a grant form ENS Lyon. L.S. is supported by the Labex ENIGMASS, initiative d'excellence.

\end{document}

%% file: defs.tex
\newcommand{\qsubrm}[2]{{#1}_{\scriptscriptstyle{\textrm{#2}}}}

\newcommand{\qsuprm}[2]{{#1}^{\scriptscriptstyle\textrm{#2}}}

\def\be{\begin{equation}}
\def\ee{\end{equation}}
\def\bea{\begin{eqnarray}}
\def\eea{\end{eqnarray}}
\def\bse{\begin{subequations}}
\def\ese{\end{subequations}}

\let\oldsqrt\sqrt
\def\sqrt{\mathpalette\DHLhksqrt}
\def\DHLhksqrt#1#2{%
\setbox0=\hbox{$#1\oldsqrt{#2\,}$}\dimen0=\ht0
\advance\dimen0-0.2\ht0
\setbox2=\hbox{\vrule height\ht0 depth -\dimen0}%
{\box0\lower0.4pt\box2}}